\documentclass[graybox, secnum]{svmult}

\usepackage{newtxtext}
\usepackage{helvet}         
\usepackage{courier}        
\usepackage{type1cm}        
\usepackage{makeidx}         
\usepackage{graphicx}        
\usepackage{multicol}        
\usepackage[bottom]{footmisc}
\usepackage{hyperref}        
\usepackage{soul}            
\hypersetup{colorlinks=true,urlcolor=blue}
\usepackage[square,numbers]{natbib}
\makeindex             
\usepackage{amsmath,amssymb}
\def\ie{i.e.,~}
\def\eg{e.g.,~}

\newcommand{\rmd}{{\rm d}}

\newcommand{\ex}{\hat{\bb{e}}_x}
\newcommand{\ey}{\hat{\bb{e}}_y}
\newcommand{\ez}{\hat{\bb{e}}_z}
\newcommand{\eb}{\hat{\bb{b}}}
\newcommand{\ek}{\hat{\bb{k}}}
\newcommand{\pD}[2]{\frac{\partial #2}{\partial #1}}

\newcommand{\Deriv}[2]{\frac{\rmd #2}{\rmd #1}}

\newcommand\bb[1]{\mbox{\boldmath{$#1$}}}
\newcommand\grad{\bb{\nabla}}
\newcommand\bcdot{\,\bb{\cdot}\,}
\newcommand\btimes{\,\bb{\times}\,}
\newcommand{\mc}[1]{\mathcal{#1}}
\newcommand{\msb}[1]{\bb{\mathsf{#1}}}
\newcommand\mfp{\lambda_{\rm mfp}}
\newcommand\pangle{\xi}

\begin{document}
\title*{Plasma physics of the intracluster medium}
\author{Matthew W.~Kunz\thanks{corresponding author}, Thomas W.~Jones, and Irina Zhuravleva}
\institute{Matthew W.~Kunz \at Department of Astrophysical Sciences, Princeton University, Princeton, NJ USA 08540 \email{mkunz@princeton.edu}
\and Thomas W.~Jones \at School of Physics and Astronomy, University of Minnesota, Minneapolis, MN USA 55455 \email{twj@umn.edu} 
\and Irina Zhuravleva \at Department of Astronomy and Astrophysics, University of Chicago, Chicago IL USA 60637 \email{zhuravleva@astro.uchicago.edu}}
%
%
\maketitle
\abstract{This Chapter provides a brief tutorial on some aspects of plasma physics that are fundamental to understanding the dynamics and energetics of the intracluster medium (ICM). The tutorial is split into two parts: one that focuses on the thermal plasma component -- its stability, viscosity, conductivity, and ability to amplify magnetic fields to dynamical strengths via turbulence and other plasma processes; and one that focuses on the non-thermal population of charged particles known as cosmic rays -- their acceleration, re-acceleration, and transport throughout the cluster volume. Observational context is woven throughout the narrative, from constraints on the strength and geometry of intracluster magnetic fields and the effective viscosity of the ICM, to examples of radio halos, radio relics, and cluster shocks that can test theories of particle acceleration. The promise of future X-ray missions to probe intracluster turbulence and discover the impact of small-scale plasma physics, coupled with sensitive, high-resolution radio observations of synchrotron-emitting plasma that reveal the properties of intracluster magnetic fields and particle-acceleration mechanisms, are likely to establish galaxy clusters as the premier cosmic laboratories for deciphering the fundamental physics of hot, dilute plasmas.}


\newpage

\section{Introduction}

This Chapter takes as its starting point the view of galaxy clusters as cosmic laboratories for deciphering the fundamental physics of hot, dilute, astrophysical plasmas. What makes this view now tenable is a recent set of (abbreviated) {\em Hitomi} and long-exposure {\em Chandra} observations of X-ray emission from the intracluster medium (ICM) within the nearby Perseus and Coma clusters, whose analyses have focused on assessing quantitatively its material properties (namely, its viscosity and thermal conductivity) alongside its dynamical state (e.g., turbulence amplitude and spectrum). Coupled with the rich history of radio observations revealing intracluster magnetic fields and populations of non-thermal (relativistic) particles, galaxy clusters are emerging at the forefront of observational plasma physics.

With this focus borne in mind, we embark on an abridged tour of plasma physics relevant to both the thermal and non-thermal components of galaxy clusters. We spend no time educating the reader on basic magnetohydrodynamics (MHD), but rather focus on what makes the ICM {\em not} MHD -- from the disparity between the particles' Larmor radii, collisional mean free paths, and macroscopic (gradient) lengthscales (\S\ref{sec:scales}), to the collisionless shock physics that governs the acceleration of relativistic charged particles (\S\ref{sec:DSA}). Observational context is spread throughout the text.

Before proceeding, we caution the reader that the purpose of this Chapter is purely pedagogical. We aspire to provide (what we believe to be) the minimal set of plasma physics required to understand and leverage contemporary and future observations of the intracluster medium and intergalactic medium, in a concise and (hopefully) accessible way. The reader will find herein neither an adequate literature review of past and contemporary efforts to understand the hydrodynamic and thermodynamic state of these plasmas, nor a complete summary of the relevant observations and their analyses. The topics emphasized here are instead a reflection of the expertise and inclinations of this Chapter's authors, and not necessarily of those belonging to the broader community of X-ray astronomers and astrophysical fluid/plasma dynamicists. {\it Caveat emptor.}

\section{Plasma physics of the thermal ICM}

\subsection{Scale hierarchy}\label{sec:scales}

The most important characteristic of the thermal ICM from the standpoint of plasma physics concerns the extensive hierarchy of spatial and temporal scales on which its dynamically important physics lives. At the very top are the cluster-gradient lengthscales, $L \sim 100~{\rm kpc}$, which characterize macroscopic variations in the ICM number density $n$ and temperature $T$ (with Boltzmann's constant $k_{\rm B}$ subsumed into its definition unless degrees Kelvin are used explicitly), and the timescale over which sonic disturbances cross them, $L/v_{\rm th,i} \sim 100~{\rm Myr}$, where $v_{\rm th,i} \doteq (2T/m_{\rm i})^{1/2} \sim 10^3~{\rm km~s}^{-1}$ is the thermal speed of ions with mass $m_{\rm i}$ at a typical mean cluster temperature of ${\sim}5~{\rm keV}$. The fact that this thermal speed is a factor of ${\sim}3$--$10$ times larger than observationally inferred large-scale turbulent velocities $U$  implies that such motions are approximately incompressible, with associated eddy-turnover times at the outer scale ${\sim}L/U$ being in the range of ${\sim}300~{\rm Myr}$ to ${\sim}1~{\rm Gyr}$. Plasma processes characterized by larger (i.e., order-unity) Mach numbers $M$, such as cluster shocks, will only be treated in this Chapter in the context of particle acceleration (\S\ref{sec:DSA}).

The next smallest scale is the Coulomb-collisional mean free path $\mfp$, which ranges from ${\sim}0.1~{\rm kpc}$ in the deep interiors of cool-core clusters up to ${\sim}10~{\rm kpc}$ in their outskirts (and throughout particularly hot, approximately isothermal clusters such as Coma). With the ratio $L/\mfp \sim 10$--$10^3$, we are safe in treating the macroscopic dynamics as fluid-like, but perhaps only marginally so -- a state to which we refer as  ``weakly collisional''. Departures from local thermodynamic equilibrium are expected to be of order $\mfp/L$ and, as we discuss in \S\ref{sec:micro}, values as small as ${\sim}1\%$ can be dynamically important. 

From $\mfp$, it's a long way down to the next dynamically important lengthscale -- the ion-Larmor radius $\rho_{\rm i} = v_{\rm th,i}/\Omega_{\rm i}$, on which thermal ions execute gyromotion about the cluster magnetic field $\bb{B}$ at the ion-Larmor frequency $\Omega_{\rm i}\doteq eB/m_{\rm i}c$ ($e$ is the elementary charge and $c$ is the speed of light). With typical observed intracluster magnetic-field strengths $B \sim 3~\mu{\rm G}$~\cite{ct02,govoni17}, corresponding to a characteristic Alfv\'{e}n speed $v_{\rm A} \doteq B/\sqrt{4\pi m_{\rm i}n} \sim 100~{\rm km~s}^{-1}$, we find $\rho_{\rm i} \sim 1~{\rm npc}$ -- fourteen orders of magnitude smaller than $L$.\footnote{For reference, the radius of Jupiter is ${\simeq}2.3~{\rm npc}$. Because of the ICM's high temperatures and enormous size, a magnetic-field strength ${\gtrsim}3\times 10^{-18}~{\rm G}$ is sufficient to make $\rho_{\rm i}/L \lesssim 1\%$.} Despite $\rho_{\rm i}$ being cosmologically infinitesimal, the fact that $\rho_{\rm i} \lll \mfp \ll L$ is hugely important for the transport properties and dynamical stability of the ICM. At the very least, it strongly makes the case for an MHD treatment of ICM dynamics. The relatively long mean free path and the fact that $v_{\rm th,i}\gg v_{\rm A}$ demand that we go further. The next three sections explain why.

\subsection{Plasma magnetization and anisotropic transport}\label{sec:transport}

Section~\ref{sec:scales} established that the ICM may be viewed as a weakly collisional, magnetized fluid. Here we discuss what this means as concerns the large-scale transport of momentum and heat. The essential point is that, with $\rho_{\rm i}\lll\lambda_{\rm mfp}\ll L$, gyrating particles are almost perfectly confined to magnetic lines of force, a property that stifles binary (Coulomb) interactions between particles gyrating about different field lines. As a result, gradients in velocity and temperature that are oriented across the local magnetic field are difficult to relax through collisional transport alone. By contrast, those gradients oriented along the local magnetic field can be readily sampled by particles whose ability to traverse a collisional mean free path is unimpeded by magnetic forces. Mathematically, one simply makes the following replacements to the diffusive fluxes of momentum ($\msb{\Pi}$) and heat ($\bb{q}$) that are seen in the more conventional Navier--Stokes or Chapman--Enskog equations~\cite{braginskii65}:
\begin{subequations}\label{eqn:flux}
\begin{eqnarray}
    \msb{\Pi} \doteq -mn\kappa_u \msb{W} &\longrightarrow &-mn\kappa_u\, \frac{3}{2} \biggl(\eb\eb - \frac{\msb{I}}{3}\biggr) \biggl(\eb\eb - \frac{\msb{I}}{3}\biggr)\bb{:}\,\msb{W} \doteq \msb{\Pi}_\parallel, \label{eqn:uflux} \\
    \bb{q} \doteq -n\kappa_T \grad T &\longrightarrow &-n\kappa_T\,\eb\eb\bcdot\grad T \doteq \bb{q}_\parallel, \label{eqn:Tflux}
\end{eqnarray}
\end{subequations}
where 
\begin{equation}
    \msb{W} \doteq \grad\bb{u} + (\grad\bb{u})^\top - \frac{2}{3}(\grad\bcdot\bb{u})\msb{I} 
\end{equation}
is the (traceless, symmetric) rate-of-strain tensor (the superscript $\top$ denotes the transpose), $\bb{u}$ is the bulk fluid velocity, $\msb{I}$ is the unit dyadic, and $\eb \doteq \bb{B}/B$ is the unit vector in the direction of the magnetic field.\footnote{The double-dot product $\msb{C}\,\bb{:}\,\msb{D}$ of two rank-2 tensors $\msb{C}$ and $\msb{D}$ is the scalar $\sum_{ij} C_{ij} D_{ij}$.} 

The diffusion coefficients $\kappa_u$ and $\kappa_T$, both proportional to $v_{\rm th}\mfp$, characterize the rate at which the diffusive exchange of particle momentum and energy seek to establish global thermodynamic equilibrium by erasing free-energy gradients in the plasma. When $\mfp$ is determined by Coulomb collisions alone, $v_{\rm th}\mfp \propto 1/\sqrt{m}$, and so ions (electrons) dominate the diffusive transfer of momentum (heat). In this case, $\kappa_u \simeq 0.48 v^2_{\rm th,i}\tau_{\rm i}$ and $\kappa_T \simeq 1.58 v^2_{\rm th,e} \tau_{\rm e}$, where
\begin{equation}
    \tau_{\rm i} \doteq \frac{3 \sqrt{m_{\rm i}} T^{3/2}_{\rm i}}{4\sqrt{\pi} n \lambda_{\rm i} e^4} \quad{\rm and}\quad \tau_{\rm e} \doteq \frac{3 \sqrt{m_{\rm e}} T^{3/2}_{\rm e}}{4\sqrt{2\pi} n \lambda_{\rm e} e^4} \tag{\theequation {a,b}}
\end{equation}
are the ion and electron collision timescales, and $\lambda_\alpha$ is the appropriate Coulomb logarithm for species $\alpha={\rm i},{\rm e}$ (the precise numerical pre-factors in the diffusivities depend on the form of the adopted collision operator). These two quantities, $\kappa_u$ and $\kappa_T$, are sometimes referred to in the astrophysical community as the ``Spitzer viscosity'' and ``Spitzer conductivity'', following Ref.~\cite{spitzer62}. Throughout the remainder of this Chapter, we do not differentiate between the ion and electron temperatures, $T_{\rm i}$ and $T_{\rm e}$, unless stated explicitly. That being said, the characteristic timescale for ion-electron temperature equilibration, $\tau_{\textrm{eq,i-e}} \doteq (m_{\rm i}/2m_{\rm e})\tau_{\rm e}$, may not always be short enough for the difference between $T_{\rm i}$ and $T_{\rm e}$ to be ignored, e.g., $\tau_{\textrm{eq,i-e}} \sim 100~{\rm Myr}$ at $n \sim 10^{-3}~{\rm cm}^{-3}$ and $T\sim 5~{\rm keV}$.

When substituted into the hydrodynamic equations for the momentum and internal energy, equation~\eqref{eqn:flux} indicates that momentum and heat diffuse only along magnetic-field lines and that only those gradients oriented along $\eb$ can be relaxed. Corrections to equation~\eqref{eqn:flux} accounting for finite-Larmor-radius effects and field-perpendicular collisional transport are, respectively, factors of ${\sim}(\rho/\mfp)$ and ${\sim}(\rho/\mfp)^2$ smaller~\cite{braginskii65}. As a result, on timescales longer than the diffusive timescales ${\sim}L^2/\kappa$, field lines tend towards becoming isotachs and isotherms. Indeed, one often encounters arguments in the ICM literature about draped magnetic fields ``insulating'' cluster cold fronts from the surrounding hot gas (see references in Chapter ``The merger dynamics of the X-ray emitting plasma in clusters of galaxies'') or protecting the coherence of cold fronts and pressure-inflated bubbles by suppressing mixing~\cite{fabian03,ds09,zuhone15}. Field-aligned conduction and viscosity are also known to change fundamentally the convective stability properties of the ICM, with the temperature gradient (rather than the entropy gradient) becoming the discriminating quantity that determines stability~\cite{balbus00,balbus01,quataert08,kunz11b,xk16}. For plasmas hosting isotropically tangled magnetic fields~\cite{skilling74,rr78,cc98,nm01,cm04,komarov14} or perhaps fields replete with magnetic mirrors~\cite{chandran99,albright01,komarov16}, the fact that heat diffuses predominantly along field lines appreciably reduces the ability of the bulk plasma to isothermalize itself conductively.

For what follows, it is useful to make a connection between this fluid-focused point of view of diffusive transport and a particle-centric point of view. Each of the diffusive fluxes in equation~\eqref{eqn:flux} correspond to particular distortions in the velocity distribution function of the particles, which are produced by large-scale gradients in the background fluid velocity and temperature and reined in by collisional mixing of particle momentum and energy. For example, consider two stationary, neighboring fluid elements, isolated from one another and in local thermodynamic equilibrium: one to the left containing plasma at a ``hot'' temperature $T_{\rm h}$, and one to the right containing plasma at a ``cold'' temperature $T_{\rm c}$. If the particles within these two elements are suddenly allowed to mix, then the particles arriving from the left will generally have higher energies than those arriving from the right, and the statistical distribution of the velocity component in the direction of the temperature gradient will be skewed somewhat in the direction of motion of these hotter particles. As a result, there will be a small (${\sim}\mfp/L$) flux of energy between the two fluid elements, approximately linear in the temperature gradient, which collisions attempt to reduce in an effort to achieve global thermal equilibrium. When particles and their binary interactions are constrained by magnetic forces, the associated velocity-space distortion in the particle distribution function is constrained to occur predominantly in the direction of the local magnetic field. The anisotropic heat transport captured by equation~\eqref{eqn:Tflux} then follows.

\subsection{Adiabatic invariance and temperature anisotropy}

This focus on statistical biases in the charged particles' motion is particularly useful for understanding the origin of the momentum flux expressed by the right-hand side of equation~\eqref{eqn:uflux}. 

Consider a single charged particle gyrating periodically about a magnetic field at the frequency $\Omega$. Suppose the field to be spatially uniform but with a strength $B$ that changes slowly in time at the rate $\omega$. If $\omega\ll\Omega$, then the magnetic moment of the particle $\mu \doteq m w^2_\perp/2B$, where $\bb{w}_\perp$ is the component of the particle's velocity oriented perpendicular to the magnetic field, is an {\em adiabatic invariant}. Namely, as the magnetic-field strength changes, the perpendicular velocity of the particle adjusts to conserve $\mu$ -- not exactly, but nearly so ($\mu$ is ``constant to all orders'' in an expansion in $\omega/\Omega$, meaning that it is conserved to exponential accuracy~\cite{kruskal58}). This may be viewed profitably as conservation of the amount of magnetic flux threading the particle's orbit, $B\pi\rho^2$ with $\rho=w_\perp/\Omega$ being the particle's gyro-radius -- a property that clearly becomes untenable when $\omega/\Omega \gtrsim 1$. If we were to have a thermally distributed collection of such particles with number density $n$, each conserving their own $\mu$, then the expectation value of $\mu$ would of course also be conserved: $\langle\mu\rangle \doteq (1/n)\langle\mu\rangle = (1/n)\langle mw^2_\perp/2B\rangle = P_\perp/nB$, where we have introduced $P_\perp \doteq \langle m w^2_\perp/2\rangle$ as the perpendicular pressure, i.e., the expectation value of the perpendicular kinetic energy of the particles. Thus, if $B$ were to increase slowly, then the perpendicular temperature $T_\perp\doteq P_\perp/n$ would also increase under $\mu$ conservation. Unless the parallel temperature $T_\parallel \doteq P_\parallel/n \doteq (1/n)\langle mw^2_\parallel\rangle$, where $w_\parallel$ is the component of the particle's velocity oriented parallel to the magnetic field, increases in the same manner -- and in general, it does not -- then {\em temperature anisotropy} arises in the plasma, with the plasma particles' energies being biased in the direction parallel ($T_\parallel > T_\perp$) or perpendicular ($T_\perp > T_\parallel$) to the local magnetic-field direction~\cite{cgl56}. 

Just how large the temperature anisotropy $\Delta T\doteq T_\perp - T_\parallel$ can become depends on a number of things, but in the simplest case the rate of change of the magnetic-field strength (which promotes $\Delta T\ne 0$ through adiabatic invariance) and the plasma collisionality (which drives $\Delta T\rightarrow 0$ by isotropizing the particles' velocity distribution function) are in control. When the collision frequency $\nu$ is much greater than the rate of change of the magnetic-field strength in some fluid element, the local temperature anisotropy can be approximated by the steady-state relation
\begin{equation}\label{eqn:DT1}
    \frac{\Delta T}{T} \approx \frac{1}{\nu} \Deriv{t}{\ln B} ,
\end{equation}
where $\rmd/\rmd t \doteq \partial/\partial t + \bb{u}\bcdot\grad$ is the time derivative measured comoving with that fluid element at the velocity $\bb{u}$. Further taking into account conservation of a second adiabatic (``bounce'') invariant\footnote{The second adiabatic invariant $\mc{J}$ is related to the periodic motion of a $\mu$-conserving particle's guiding center as it bounces between two ends of a slowly lengthening or shortening magnetic bottle. For the purposes of this Chapter, it suffices to say that $\mc{J}$ conservation makes the field-parallel velocity of the particle $w_\parallel$ inversely proportional to the length of the bottle $\ell$, the latter of which can be related in an MHD fluid to the ratio $B/n$~\cite{kulsrud83}. As a result, $\mc{J}$ conservation for a thermally distributed population of particles is often expressed as $T_\parallel \propto (B/n)^2$.} provides the following update to equation~\eqref{eqn:DT1}:
\begin{equation}\label{eqn:DT2}
    \frac{\Delta T}{T} = \frac{1}{\nu} \Deriv{t}{} \ln \frac{B^3}{n^2} .
\end{equation}
Provided one knows how the magnetic-field strength $B$ and density $n$ evolve, as well as how to compute the collision frequency $\nu$, equation~\eqref{eqn:DT2} offers a means of estimating deviations from local thermodynamic equilibrium. The ideal-MHD induction equation and mass continuity equation provide us with the former:
\begin{eqnarray}
    \pD{t}{\bb{B}} = \grad\btimes(\bb{u}\btimes\bb{B}) &\Longrightarrow &\Deriv{t}{\ln B} = (\eb\eb-\msb{I})\,\bb{:}\,\grad\bb{u} \label{eqn:lnB}, \\
    \pD{t}{n} = -\grad\bcdot(n\bb{u}) &\Longrightarrow &\Deriv{t}{\ln n} = -\grad\bcdot\bb{u} .
\end{eqnarray}
Substituting these expressions back into equation~\eqref{eqn:DT2} gives
\begin{equation}\label{eqn:DT3} 
     \frac{\Delta T}{T} = \frac{3}{\nu} \left(\eb\eb - \frac{\msb{I}}{3}\right)\bb{:}\,\grad\bb{u} = \frac{3}{2\nu} \left(\eb\eb - \frac{\msb{I}}{3}\right)\bb{:}\,\msb{W} .
\end{equation}
The form of the final equality here should look familiar from \S\ref{sec:transport}. Namely, if we associate the collision frequency $\nu$ with ${\simeq}(0.96\tau_{\rm i})^{-1}$, then equation~\eqref{eqn:uflux} becomes
\begin{equation}
    \msb{\Pi} = -\left(\eb\eb-\frac{\msb{I}}{3}\right) \Delta P,
\end{equation}
where $\Delta P = n \Delta T = P_\perp - P_\parallel$ is the {\em pressure anisotropy}. In words, the viscous stress experienced by a fluid element in a weakly collisional, magnetized plasma is directly related to the velocity-space anisotropy of the constituent charged particles, which is biased with respect to the magnetic-field direction by adiabatic invariance and relaxed by isotropizing collisions.

This closure for the viscous stress, derived by Braginskii~\cite{braginskii65}, offers a means of estimating deviations from local thermodynamic equilibrium in the ICM. With equation~\eqref{eqn:DT3} giving $\Delta T/T \sim (\mfp/L) (u/v_{\rm th,i}) \sim 10^{-2}$, one may be forgiven for thinking none of this matters a great deal. However, the main competition of the associated viscous stress in the fluid momentum equation is the magnetic (``Maxwell'') stress, and so the proper comparison to make is not $\Delta T/T$ versus unity, but rather $\Delta P$ versus $B^2/4\pi$. With observationally inferred values of $\beta\doteq 8\pi nT/B^2 \sim 10^2$ and turbulent bulk velocities $u \sim v_{\rm A}$, it seems to be the case that the turbulent kinetic energy, magnetic energy, and pressure anisotropy are all in approximate equipartition.\footnote{The contribution to the total pressure anisotropy from the electrons is expected to be ${\sim}\sqrt{m_{\rm e}/m_{\rm i}}$ times smaller in the weakly collisional ICM, and thus likely to be of little consequence.} There are several macro-scale consequences of such a dynamically important viscous stress, from its ability to nullify magnetic tension and thereby interrupt the propagation of nonlinear Alfv\'en waves~\cite{squire16,squire17}, to its impact on the turbulent amplification of weak magnetic fields~\cite{stonge20}. Some of these consequences are highlighted in \S\ref{sec:example_dynamo}. Of perhaps more immediate concern, however, is that there is a crucially important {\em microscale} consequence of having $\Delta T/T \sim 1/\beta$ -- one which is routinely measured by {\em in situ} spacecraft in the solar wind \cite{kasper02,hellinger06,bale09,chen16}, and to which we now turn.

\subsection{Kinetic micro-instabilities and their impact on transport}\label{sec:micro}

We have just seen that even very small temperature anisotropies produce a viscous stress that, in the high-$\beta$ ICM, can rival the Maxwell stress. Such a situation is well known in the plasma-physics community to be kinetically unstable to a variety of rapidly growing, Larmor-scale fluctuations, with names such as ``firehose'' and ``mirror''. This Chapter is not the place for a full exposition on such ``micro-instabilities'', but there are nevertheless a few important implications for ICM physics that ought to be listed, explained, and, ultimately, heeded. All of them concern the manner in which these instabilities interact with plasma particles to supplant particle-particle collisions with wave-particle interactions as the dominant contribution to the plasma's viscosity and thermal conductivity. While this is an ongoing area of research, there seem to be a few reliable take-away points, which seem to be consistent with current observational constraints and which we believe should become part of the vernacular of ICM astrophysicists.

When the temperature anisotropy satisfies $\Delta T/T \lesssim -2/\beta$, the restoring tension force responsible for the propagation of Alfv\'enic fluctuations is undermined by an opposing viscous stress, and those fluctuations no longer propagate but rather grow exponentially in what is known as the firehose instability~\cite{rosenbluth56,chandrasekhar58,parker58,hm00}.\footnote{We have used $\lesssim$ rather than $\le$ in the firehose instability criterion, as various finite-Larmor-radius and Landau-resonant effects collaborate to reduce slightly the numerical pre-factor $2$ in a way not amenable to providing a single number or even simple explanation~(AFA~Bott, private communication).} The fluctuations that grow the fastest reside near the ion-Larmor scale and have growth rates a substantial fraction of the ion-Larmor frequency. Mathematically, the Maxwell stress on the conducting fluid caused by a magnetic perturbation $\delta\bb{B}$ to a mean field $\bb{B}$, {\em viz.}~$(\bb{B}\bcdot\grad)\delta\bb{B}$, is effectively replaced by $(1 + 4\pi\Delta P/B^2)(\bb{B}\bcdot\grad)\delta\bb{B}$, whose sign can flip if the above instability criterion is satisfied. In this case, it is energetically favorable to amplify magnetic perturbations at the expense of the temperature anisotropy, increasingly so at smaller scales (down to ${\sim}\rho_i$). The practical consequence is that ions eventually pitch-angle scatter off these firehose fluctuations, thereby restoring the local temperature anisotropy to marginally unstable values ({\em viz.}, $\Delta T/T \approx -2/\beta$) and reducing effectively the plasma viscosity below its Coulombic value -- an effect long suspected but only recently understood in detail theoretically \cite{kunz14,ht15,msk16}. This reduction is quantified later in this subsection.

When the temperature anisotropy instead satisfies $\Delta T/T \gtrsim 1/\beta$, the plasma is also susceptible to rapidly growing, ion-Larmor-scale magnetic perturbations -- this time polarized compressively, meaning that they involve (anti-)correlated fluctuations in density and magnetic-field strength, rather than Alfv\'enically. This is the mirror instability~\cite{barnes66,hasegawa69,hellinger07}, essentially a magnetic bottle that has been destabilized by an excess of perpendicular pressure over magnetic pressure that is reinforced as the bottle grows in amplitude. (On a more technical level: Landau-resonant guiding centers extract energy from the magnetic mirror force near the ends of the bottle, increasingly flooding its interior with particles having preferentially large perpendicular energies, inflating the ballooned magnetic-field lines further, and thereby reinforcing the mirror force~\cite{sk93}.) Once such ion-Larmor-scale magnetic mirrors grow to have amplitudes $\delta B/B \sim 0.3$, they join the firehose instability in pitch-angle scattering ions, thereby regulating the temperature anisotropy down to marginally unstable values and decreasing the effective viscosity of the plasma~\cite{kunz14,riquelme15,ht15}.

From a macroscopic point of view, what this all means is that the bulk ICM is likely polluted with microscale magnetic fluctuations that scatter ions and thereby provide an anomalous ``effective'' viscosity that supplants (or at least supplements) the more customary Coulombic one~\cite{scheko05,lyutikov07,kunz11a}. Given the various bits of observational evidence for suppressed viscosity in the ICM (\S\ref{sec:obsvisc}), it is worth quantifying this effective viscosity based on knowledge gleaned from recent kinetic, particle-in-cell simulations. The first step in doing so is to adopt the firehose/mirror scattering frequency measured directly in these simulations, $\nu_{\rm eff} \sim \beta|\eb\eb\bb{:}\grad\bb{u}|$~\cite{kunz14,msk16}. From equation~\eqref{eqn:DT2}, it is clear that such a collisionality is sufficient to maintain $|\Delta T/T| \sim 1/\beta$, that is, to restrain the temperature anisotropy to be bounded by the thresholds for the firehose and mirror instabilities. The next step is to determine the size of the field-parallel rate of strain, $|\eb\eb\bb{:}\grad\bb{u}|$. For that, we assume a customary Kolmogorov turbulence cascade in which the velocity increment on scale $\ell$ satisfies $u_\ell \propto \ell^{1/3}$~\cite{kolmogorov41}. The rate of strain $|\grad\bb{u}|\sim u_\ell/\ell \propto \ell^{-2/3}$ is then dominated by those eddies residing near the smallest available scale, namely, that set by viscosity. At this {\it viscous scale} $\ell_{\rm visc}$, the nonlinearity $\bb{u}\bcdot\grad\bb{u}$ responsible for cascading the turbulent energy to smaller scales becomes comparable to the viscous force $\kappa_u\nabla^2\bb{u}$ at which that energy is dissipated. Adopting $u_\ell \sim U (\ell/L)^{1/3}$ then provides $\ell_{\rm visc} \sim L\,{\rm Re}^{-3/4}$, where ${\rm Re}\doteq UL/\kappa_u$ is the Reynolds number, at which $|\grad\bb{u}| \sim (U/L) {\rm Re}^{1/2}$. For a magnetized plasma, we simply replace ${\rm Re}$ by ${\rm Re}_\parallel$, the subscript indicating that the field-parallel component of the viscosity is to be used. When that parallel viscosity is determined by the firehose/mirror collisionality, the Kolmgorov scaling for the viscous-scale rate of strain results in an implicit equation for $\nu_{\rm eff}$ -- one whose solution provides $\nu_{\rm eff} \sim (U/L) (U/v_{\rm A})^{2} \beta$. This scaling in turn implies the following effective field-parallel mean free path, parallel Reynolds number, and parallel viscous scale:
\begin{subequations}\label{eqn:predscales}
\begin{equation}
    \lambda_{\rm \parallel mfp, eff} \sim L\,\biggl(\frac{U}{v_{\rm A}}\biggr)^{-3}\beta^{-1/2}, \quad {\rm Re}_{\parallel\rm ,eff} \sim \biggl(\frac{U}{v_{\rm A}}\biggr)^4, \quad \ell_{\rm \parallel visc,eff} \sim L \, \biggl(\frac{U}{v_{\rm A}}\biggr)^{-3}, \tag{\theequation {a,b,c}}
\end{equation}
\end{subequations}
respectively~\cite{stonge20}. Note that the effective viscous scale is also the Alfv\'en scale, on which $u_\ell\sim v_{\rm A}$. If the total magnetic energy turns out to be some fraction $f_{B}$ of the bulk kinetic energy -- a typical outcome in theory and simulation of turbulent dynamo (\S\ref{sec:example_dynamo}) -- then $\lambda_{\parallel\rm mfp,eff} \sim f^2_{B} M$ and ${\rm Re}_{\parallel\rm ,eff} \sim f^{-2}_{B}$.

\begin{figure}
\includegraphics[width=\linewidth]{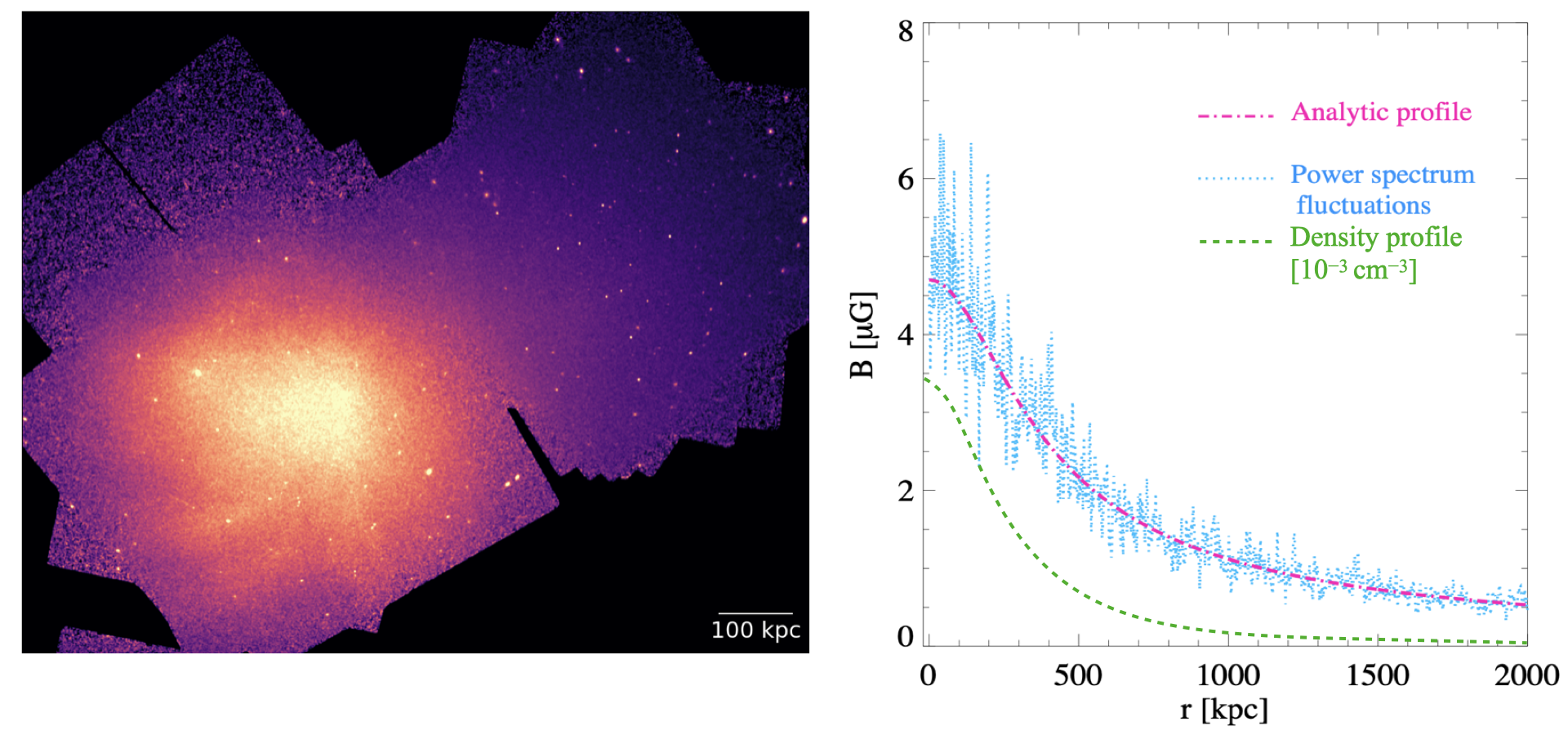}
\caption{{\bf Left:} X-ray image (0.5--3.5 keV energy band) of the Coma cluster observed with {\it Chandra}. Yellow/purple color corresponds to the highest/lowest X-ray surface brightness. For display purposes, the image is lightly smoothed. Adapted from Ref.~\cite{zhuravleva19}. {\bf Right:} Best-fitting radial ($r$) profile of magnetic-field strength measured through an analysis of RMs around seven radio sources in the Coma cluster observed by the VLA (magenta). The blue line indicates simulated power spectrum fluctuations on the profile. A fit to the gas density profile is shown in green. Adapted from Ref.~\cite{bonafede10}.}
\label{fig:coma_image_mf}
\end{figure}

Blindly promoting the ``$\sim$'' in equation~\eqref{eqn:predscales} to ``$\approx$'', and normalizing $U$, $B$, $n$, and $T$ using parameters similar to those observed in the Coma cluster~\cite{bonafede10,zhuravleva19} (see figure~\ref{fig:coma_image_mf}), yields the following expressions for the field-parallel effective mean free path and viscous scale:
\begin{align}
    \frac{\lambda_{\rm \parallel mfp,eff}}{L} &\approx 0.04\,\biggl(\frac{U}{200~{\rm km~s}^{-1}}\biggr)^{-3}\biggl(\frac{B}{2~\mu{\rm G}}\biggr)^4 \biggl(\frac{n}{10^{-3}~{\rm cm}^{-3}}\biggr)^{-2} \biggl(\frac{T}{10^8~{\rm K}}\biggr)^{-1/2}, \label{eqn:mfpeff}\\*
    \frac{\ell_{\rm \parallel visc,eff}}{L} &\approx 0.3\,\biggl(\frac{U}{200~{\rm km~s}^{-1}}\biggr)^{-3} \biggl(\frac{B}{2~\mu{\rm G}}\biggr)^{3} \biggl(\frac{n}{10^{-3}~{\rm cm}^{-3}}\biggr)^{-3/2} .
\end{align}
%
%
Choosing $L=100~{\rm kpc}$ and comparing equation~\eqref{eqn:mfpeff} to its Coulomb-collisional counterpart, $\lambda_{\rm mfp,i}=v_{\rm th,i}\tau_{\rm i}$, we find that
\begin{equation}\label{eqn:viscsupp}
    \frac{\lambda_{\rm \parallel mfp,eff}}{\lambda_{\rm mfp,i}} \approx 0.05\,\biggl(\frac{U}{200~{\rm km~s}^{-1}}\biggr)^{-3}\biggl(\frac{B}{2~\mu{\rm G}}\biggr)^4 \biggl(\frac{n}{10^{-3}~{\rm cm}^{-3}}\biggr)^{-1} \biggl(\frac{T}{10^8~{\rm K}}\biggr)^{-5/2} .
\end{equation}
Note that an isotropically tangled field on sub-viscous scales would have the anisotropic viscous stress, proportional to $\eb\eb\eb\eb$, act as though $\mfp = \lambda_{\rm \parallel mfp}/5$~\cite{mk02}, thereby further reducing the above suppression factor to ${\approx}0.01$. While these values are consistent with current constraints on the suppression of Spitzer viscosity in the Coma cluster, as inferred from the power spectrum of de-projected density fluctuations~\cite{zhuravleva19} (see \S\ref{sec:obsvisc}), there are important order-unity coefficients that have yet to be sorted out. An additional complication is that Alfv\'enic turbulence subject to Braginskii viscosity may be capable of self-organizing to minimize the projection of $\grad\bb{u}$ onto the local magnetic-field-direction tensor $\eb\eb$, thereby avoiding strong parallel-viscous dissipation in what Ref.~\cite{squire19} refers to as ``magneto-immutability''.

Given all the attention paid in recent years to various structures and turbulent velocities seen in X-ray emission from the Perseus cluster, it seems the most pressing observation to perform there from the standpoint of this subsection is a measurement of the magnetic-field strength. With {\em Hitomi} observations providing a line-of-sight velocity dispersion of ${\approx}164~{\rm km~s}^{-1}$ at a radial offset of ${\sim}50~{\rm kpc}$ in Perseus~\cite{hitomi}, where the local $n \approx 0.02~{\rm cm}^{-3}$~\cite{zhuravleva15}, the implied equipartition magnetic-field strength ($f_B=1$) is ${\approx}10~\mu{\rm G}$; the actual field strength is likely to be less than this value.\footnote{The only reported detection (to our knowledge) of a rotation measure (RM; see \S\ref{sec:obsB}) in Perseus~\cite{taylor06} implies a magnetic-field strength ${\sim}20~\mu{\rm G}$ in the highest-density part of the cluster ($n \approx 0.3~{\rm cm}^{-3}$) near the compact radio source 3C~84. Simply for the sake of argument, if one were to adopt the scaling $B\propto n^{0.47}$ previously obtained from a least-squares fit to  field strengths inferred from RMs in the inner regions of nine nearby clusters~\cite{govoni17} and use it to extrapolate this ${\sim}20~\mu{\rm G}$ field to lower densities, then one would obtain $B\sim 5~\mu{\rm G}$ at $n\approx 0.02~{\rm cm}^{-3}$ in Perseus. A stronger scaling with density, for example the $(\rmd\ln B/\rmd\ln n) \approx 1$ scaling inferred from radial RM profiles in some clusters~\cite{vacca12,govoni17,stuardi21}, would of course imply an even smaller value for $B$ in the bulk ICM.} Using equation~\eqref{eqn:viscsupp} with $T\approx 4~{\rm keV}$ and $L\approx 20~{\rm kpc}$ -- the latter being comparable to the scales of the largest AGN-inflated bubbles in the field of observation -- a field strength ${\lesssim}5~\mu{\rm G}$ (corresponding to $f_B \lesssim 0.5$) would push $\lambda_{\parallel\rm mfp,eff}$ below the Coulomb mean free path of ${\approx}0.2~{\rm kpc}$. Thus, in Perseus the collisionality induced by kinetic instabilities is likely to be at least as important as the Coulomb collisionality in determining the viscosity.

Similar kinetic micro-instabilities are known to be triggered by, and subsequently to regulate, the conductive flow of heat along magnetic-field lines. One particularly notable example is the heat-flux-driven whistler instability~\cite{levinsoneichler92,pistinnereichler98}, in which transverse perturbations to the magnetic field grow on electron-Larmor scales to scatter the heat-flux-carrying electrons and thereby reduce the heat flux to marginally unstable values. Recent particle-in-cell simulations \cite{robergclark16,robergclark18,komarov18} have demonstrated that a  whistler-regulated heat flux is roughly independent of the temperature gradient and scales as the inverse electron plasma beta, $\beta^{-1}_{\rm e}$, as originally predicted by Ref.~\cite{pistinnereichler98}. Ref.~\cite{komarov18} demonstrated that the steady-state magnetic energy of the fluctuations satisfies $(\delta B/B)^2 \sim \beta_{\rm e}(\rho_{\rm e}/L_T)$, an extremely small level in the ICM but one that is sufficient to pitch-angle scatter electrons at a rate $\nu \sim \Omega_{\rm e} (\delta B/B)^2$ large enough to maintain a marginally unstable $q_\parallel \sim n T_{\rm e} v_{\rm th,e} \beta^{-1}_{\rm e}$. Those authors modeled their results using a heat-flux suppression factor ${\approx}3\beta^{-1}_{\rm e}(L_T/\lambda_{\rm mfp,e}+4)$, which for $\beta_{\rm e} \sim 100$ yields ${\sim}1/4$ on ${\sim}100~{\rm kpc}$ scales and a maximum suppression factor of ${\sim}0.1$ when $L_T\ll\lambda_{\rm mfp,e}$ (see also Ref.~\cite{drake21}). Combined with the additional suppression of heat conduction by a factor of ${\approx}1/3$--$1/5$ due to the trapping of particles in ion-Larmor-scale magnetic bottles grown by the mirror instability \cite{komarov16}, it is reasonable to expect a parallel heat flux on the order of a several percent of the full Spitzer value in the more diffuse, high-$\beta$ portions of the ICM.

\begin{figure}[t]
\includegraphics[width=\linewidth]{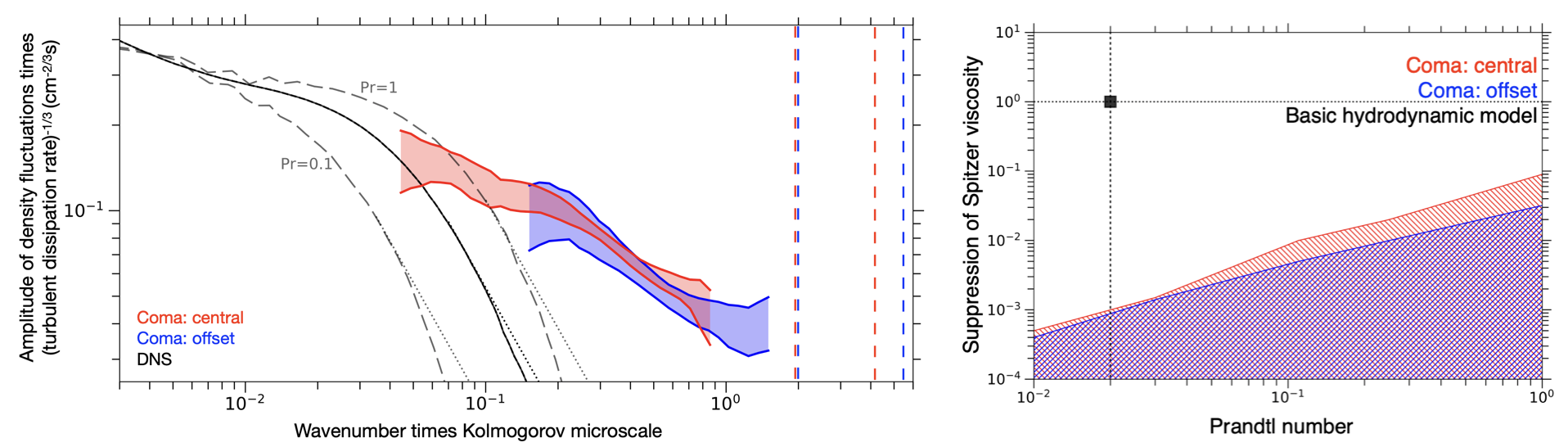}
\caption{Constraining effective viscosity in the bulk intergalactic plasma of the Coma cluster. {\bf Left:} The amplitude of density fluctuations as a function of a wavenumber measured in the central (red) and offset (blue) regions shown in figure~\ref{fig:coma_image_mf}. The scale-dependent amplitudes of velocity and density from hydrodynamic numerical simulations are shown by the black and grey curves, respectively. The shape of the density spectrum depends on the thermal Prandtl number ${\rm Pr}$, which parameterizes the strength of thermal conduction; two representative cases, ${\rm Pr}=1$ and ${\rm Pr}=0.1$, are shown. Dotted lines indicate how the spectral slopes are modified by the observational method of measuring density fluctuations amplitudes. The dashed vertical lines indicate variations in the Coulomb mean free path in the central and offset regions. The slopes of the observed spectra are inconsistent with an exponential cutoff on small scales, implying that the effective isotropic viscosity is orders of magnitude smaller than predicted by hydrodynamic models with Coulomb collision rates. {\bf Right:} Required suppression of Spitzer viscosity vs.~thermal Prandtl number. Parameters within the hatched regions provide consistency between the observed power spectra and predictions from the simulations. The black point corresponds to the purely hydrodynamic case with Coulomb collisions. Regardless of ${\rm Pr}$, the effective viscosity in the bulk ICM is suppressed by a factor of at least 10. Adapted from Ref.~\cite{zhuravleva19}.}
\label{fig:coma_viscosity}
\end{figure}

\subsection{Example: Suppressed viscosity in the Coma cluster}\label{sec:obsvisc}

Thanks to its proximity and brightness, the Coma cluster spans more than one degree in projected size on the sky. As a result, deep {\it Chandra} observations are able to resolve fluctuations in X-ray emission on scales down to those approaching the Coulomb mean free path. Recently,  Ref.~\cite{zhuravleva19} used such observations of two separate regions of the Coma cluster -- a ``central'' region focused on the innermost ${\sim}500~{\rm kpc}$ and an ``offset region'' a few hundred kpc away from the center -- to calculate the power spectrum of ICM density fluctuations spanning more than a decade in scale; see the left panel of figure~\ref{fig:coma_viscosity}. By assuming that these fluctuations are passively advected by the velocity field -- a reasonable assumption if the motions are subsonic and driven on large scales -- one can use them as an indirect probe of the turbulent velocity spectrum~\cite{gaspari14,zhuravleva14b}. If such a velocity field were subject to plasma viscosity at the Spitzer level, one would observe an exponential cutoff of the power spectrum well above the viscous scale $\ell_{\rm visc}$ (${\sim}50~{\rm kpc}$ in the offset region), as indicated by the black and grey lines obtained from direct numerical simulations of hydrodynamic turbulence. However, the observed spectra from the central (red) and offset (blue) regions follow an approximately Kolmogorov slope to smaller scales with no clear indication of an exponential cutoff, the implication being that the plasma viscosity is markedly suppressed. By comparing these observational results to the numerical simulations, one can estimate the factor by which the viscosity is suppressed; the required suppression factors are indicated by the hatched regions in the right panel of figure~\ref{fig:coma_viscosity}. It appears that the Spitzer viscosity must be suppressed by at least a factor of 10 (if not more), regardless of the level of thermal conduction in the ICM (parameterized through the thermal Prandtl number on the abscissa). This suppression factor is consistent with the arguments made in \S\ref{sec:micro} (cf.~equation~\eqref{eqn:viscsupp}), and may be taken as tentative evidence for an enhanced collisionality caused by kinetic micro-instabilities.

\subsection{Anisotropic viscosity and turbulent amplification of cluster magnetic fields}\label{sec:example_dynamo}

Magnetic-field strengths in our Galaxy and in present-day galaxy clusters are generally measured in $\mu{\rm G}$. Such values may be understood in terms of approximate equipartition with the fluid motions: in both our Galaxy and the ICM, the Alfv\'{e}n speeds implied by the observationally inferred densities and magnetic-field strengths are comparable to the observed turbulent velocities. It is then natural to attribute the amplification and sustenance of (at least the random component of) the interstellar and intracluster magnetic fields to the turbulent (or ``fluctuation'') dynamo, by which a succession of random velocity shears stretch the field and lead on average to its exponential growth~\cite{batchelor50,zeldovich84,cg95}. Similar field strengths have been inferred observationally in galaxies up to redshift $z\sim 2$~\cite{widrow02,bernet08} and from diffuse radio emission in clusters at $z\sim 0.7$~\cite{digenaro20}, indicating that such a dynamo, if active, must succeed on a cosmologically short timescale.

In this section we provide some examples of how the properties of intracluster and intergalactic magnetic fields are observationally inferred and how such fields might be created through plasma processes. The goal is to provide some relatively basic pre-requisites for understanding contemporary discussions of cosmic magnetogenesis. This summary is by no means exhaustive, and the reader would do well by consulting, e.g., Ref.~\cite{ct02} for an excellent observational overview of cluster magnetic fields and Ref.~\cite{rincon19} for a recent and authoritative review of dynamo theories. Furthermore, no attention is given here to exotic cosmological processes capable of generating primordial (``seed'') magnetic fields; for that, we point the interested reader to Refs~\cite{dn13,subramanian16}. Here, we merely summarize two important and popular plasma processes for generating seed magnetic fields -- the Biermann battery and the Weibel instability -- and then move on to explain how turbulence can amplify such seed fields to dynamical strengths. We close by highlighting some recent work on turbulent dynamo that moves beyond MHD and illustrates some of the physics presented in \S\S\ref{sec:transport}--\ref{sec:micro}. 

\subsubsection{Observational constraints on ICM/IGM magnetic fields}\label{sec:obsB}

Unfortunately, relatively little is known directly about the geometry and strength of intracluster and intergalactic magnetic fields. A few examples were given in \S\ref{sec:transport} in which intracluster magnetic fields had been invoked to explain the structural integrity of certain X-ray emission features. Along similar lines, observations of H$\alpha$ emission in some clusters that reveal the presence of colder gas structured into long, coherent filaments have been taken as evidence for dynamically important intracluster fields, because of the filaments' apparent stability against tidal shear and dissipation into the surrounding hot gas (e.g., \cite{fabian08}). If these filaments were produced in the wakes of buoyantly rising bubbles, Ref.~\cite{churazov13} argues that the advected magnetic-field lines would be anti-parallel and gradually forced together, ultimately leading to their reconnection and to resistive heating capable of powering the filaments' luminosity.

A much more direct way of determining the properties of intracluster magnetic fields relies on synchrotron polarimetry of the radio lobes of active galactic nuclei in the cores of galaxy clusters. While this approach returns what are perhaps the most reliable constraints on the strengths and geometries of intracluster magnetic fields -- generally ${\sim}1$--$10~\mu{\rm G}$, fluctuating on scales ${\sim}1$--$10~{\rm kpc}$ and decreasing radially outwards~\cite{murgia04,ve05,bonafede10,govoni17} -- the technique is not without its limitations, least of which is its need for several polarized radio sources favorably situated throughout the cluster (most strong radio sources embedded in clusters are located at the center). The method leverages the effect of {\em Faraday rotation}, in which linearly polarized light passes through a magnetized (and thus birefringent) medium and has its plane of polarization rotated because of the different indices of refraction for left and right circularly polarized radiation. The amount of rotation, $\Delta\phi = \lambda^2\times{\rm RM}$, is dependent upon the wavelength $\lambda$ of the radiation and the rotation measure
\begin{equation}\label{eqn:RM}
    {\rm RM}\propto \int \rmd\bb{\ell}\bcdot\bb{B}(\bb{\ell}) n_{\rm e}(\bb{\ell}) ,
\end{equation}
where $n_{\rm e}$ is the thermal electron number density (customarily determined from X-ray observations) and the integral is taken over the path length $\bb{\ell}$ of light propagation. With $n_{\rm e}$ measured in ${\rm cm}^{-3}$, $B$ in $\mu{\rm G}$, and $\rmd\ell$ in ${\rm kpc}$, the proportionality coefficient in equation~\eqref{eqn:RM} is ${\simeq}812~{\rm rad~m}^{-2}$. The dependence of RM on the wavelength allows one to obtain $\Delta\phi$ without knowing the initial polarization angle; note that only the line-of-sight component of the magnetic field can be obtained. Thus far, the best-studied cluster using this technique is the Coma cluster, highlighted in \S\ref{sec:micro} and figure~\ref{fig:coma_image_mf}, whose central magnetic-field strength was constrained to be ${\simeq}3.9$--$ 5.4~\mu{\rm G}$ to within $1\sigma$~\cite{bonafede10}. Compendia of magnetic-field strengths inferred from RMs in other clusters may be found in table~6 of Ref.~\cite{stuardi21} and table~1 of Ref.~\cite{kunz11a}. While almost all RMs are consistent with values of $\beta\sim 100$ in galaxy clusters, complications in interpreting RMs range from dealing with patchy and irregular RM distributions to ensuring that interactions between the radio source and the local ICM do not affect the measured RM (see the Introduction of Ref.~\cite{johnson20} for a recent summary of limitations).

Another way of estimating ICM magnetic-field strengths from synchrotron emission relies on equipartition arguments linking the field energy to the energy in the relativistic particles powering a given radio halo. The argument runs as follows~\cite{burbidge59,beck05}. The total energy of a synchrotron source is a sum of energy in relativistic electrons and protons and in magnetic fields, $U_{\rm tot}=U_{\rm e}+U_{\rm p}+U_{B}$. The magnetic energy is $U_{B}=(B^2/8\pi) \Phi V$, where $V$ is the volume of the synchrotron source and $\Phi$ is the filling factor of the field within $V$. It is a common assumption that the field and particles occupy the entire volume, i.e., $\Phi=1$. The energy in protons is assumed to be related to the energy in electrons via $U_{\rm p}=kU_{\rm e}$, with the proportionality constant $k$ depending on the acceleration mechanism of electrons (and is thus very uncertain). Finally, another assumption must be made about the energy distribution between the particles and fields. It is convenient to represent $U_{\rm tot}$ by its minimum value, which is obtained when the contributions of relativistic particles and the magnetic field are approximately equal (the so-called equipartition assumption), {\em viz.}~$U_{B}=\frac{3}{4}(1+k)U_{\rm e}$. Denoting the synchrotron luminosity of the radio source by $L_{\rm S}$, these arguments lead to the following estimate of the field strength:
\begin{equation}
    B\sim \biggl[\frac{(1+k)L_{\rm S}}{\Phi V} \biggr]^{2/7}.
\end{equation}
(The weak $2/7$ scaling is fortunate given all the assumptions involved.) Following this method, the estimated equipartition fields from observed synchrotron emission of radio halos are ${\sim}0.1$--$1~\mu{\rm G}$~\cite{feretti12}.

Synchrotron-emitting relativistic electrons within the ICM (discussed in \S\ref{sec:CR}) can also scatter photons from the Cosmic Microwave Background (CMB) to X-ray energies through inverse-Compton (IC) scattering. This produces a hard tail above the Bremsstrahlung continuum in spectra of galaxy clusters~\cite{rephaeli79}. The ratio of fluxes of the inverse-scattered CMB photons and synchrotron radiation from the same relativistic electrons, $F_{\rm IC}/F_{\rm S}$, scales with the volume-averaged magnetic field strength as $F_{\rm IC}/F_{\rm S}=U_{\rm CMB}/(B^2/8\pi)$, where $U_{\rm CMB}$ is the energy density of CMB photons. By measuring these fluxes, one could constrain the field strength. So far, only upper limits on the IC X-ray flux have been robustly obtained, providing only lower limits on the ICM magnetic field strength (e.g.,~\cite{sugawara09,finoguenov10}).

\subsubsection{Plasma theory basics for seed-field generation: Biermann and Weibel}

The Biermann battery~\cite{biermann50} and Weibel instability~\cite{weibel59} are two plasma processes that are often implicated when positing an astrophysical origin of primordial magnetic fields that is not reliant upon stellar sources. The former, presumed to act during large-scale structure formation in the early Universe in cosmological accretion shocks, \mbox{(re-)ionization} fronts, and/or cosmological linear over-densities (e.g.,~\cite{ps89,subramanian94,kulsrud97,ryu98,gnedin00,nn13}), is thought to be capable of generating large-scale fields with strengths ${\sim}10^{-25}$--${10}^{-18}~{\rm G}$. Its mechanism can be traced back to the disparate masses of ions and electrons, combined with the tendency for astrophysical plasmas to be quasi-neutral, that is, to have equal numbers of positive and negative charges on scales much larger than the Debye length $\lambda_{\rm D} \simeq 234\, (T_{\rm keV}/n_{\rm cm^{-3}})^{1/2} ~{\rm m}$. Because $m_{\rm e} \ll m_{\rm i}$, any mass-independent force accelerates the electrons much more so than the ions; in this case, an electric field $\bb{E}$ must arise to couple the ions and electrons and preserve quasi-neutrality. In the Biermann battery, this force is provided by pressure gradients in the plasma that are misaligned with the plasma's density gradients. This misalignment necessitates a ``thermo-electric field'' that has a curl, which from Faraday's law produces a magnetic field. (The effect is analogous to baroclinic forcing of vorticity, in which the lighter parts of a fluid are accelerated more by a transverse pressure gradient than are the heavier parts, leading to rotation about the fluid's local center of mass and thus vorticity production. Indeed, the Biermann battery can be thought of as a consequence of baroclinic vorticity production in the electron fluid acting in concert with Lenz's law.) Mathematically, an expression for the thermo-electric field caused by the differential acceleration of electrons and ions may be obtained from the force equation for the electron species,
\begin{equation}
    m_{\rm e} n_{\rm e} \Deriv{t}{\bb{u}_{\rm e}} = -\grad\bcdot\msb{P}_{\rm e} - e n_{\rm e} \left( \bb{E} + \frac{\bb{u}_{\rm e}}{c}\btimes\bb{B} \right) + \bb{R}_{\rm e} ,
\end{equation}
after using the smallness of the electron mass to drop the left-hand side, supposing $\bb{B}=0$ initially to remove the $\bb{u}_{\rm e}\btimes\bb{B}$ motional electric field, and neglecting any collisional friction force $\bb{R}_{\rm e}$ affecting the electron fluid. Taking the electron pressure tensor to be isotropic for the sake of simplicity, $\msb{P}_{\rm e} = P_{\rm e}\msb{I} = n_{\rm e} T_{\rm e} \msb{I}$, we obtain
\begin{equation}
    e\bb{E} \simeq - \frac{1}{n_{\rm e}} \grad P_{\rm e} = - \grad T_{\rm e} - T_{\rm e} \grad\ln n_{\rm e} .
\end{equation}
The first term on the right-hand side vanishes when $\bb{E}$ is substituted into Faraday's law, leaving
\begin{equation}\label{eqn:biermann}
    \pD{t}{\bb{B}} = -c\grad\btimes\bb{E} = \frac{c}{e} \, \grad T_{\rm e}\btimes \grad\ln n_{\rm e} .
\end{equation}
If the gradients of the electron temperature and density are not precisely aligned, the induced electric field is not electrostatic and so there is a path around which there is a potential drop, resulting in a current and thus a magnetic field. With $|\grad\ln T_{\rm e}| \sim |\grad\ln n_{\rm e}| \sim L^{-1}$ and sufficient misalignment, equation~\eqref{eqn:biermann} implies secular-in-time growth of the field strength, $\Omega_{\rm e} \sim (v_{\rm th,e}/L)^2 t$. Provided the misalignment persists, this growth carries on until the ion-Larmor radius $\rho_{\rm i} \lesssim ML$, after which the motional electric field associated with the Mach number $M$ becomes dominant over the thermo-electric field. For $M \sim 1$, $T_{\rm e} \sim 1~{\rm keV}$, and $L \sim 1~{\rm Mpc}$, this criterion corresponds to a field strength ${\gtrsim}10^{-21}~{\rm G}$.

Much stronger fields can be produced by the Weibel instability, albeit on microscopic plasma-inertial scales. This process relies on velocity-space anisotropy in the charged-particle distribution function -- associated perhaps with counter-streaming beams or interpenetrating flows~\cite{ss03,lazar09}, collisionless shocks~\cite{ml99,spitkovsky08,medvedev06,kt08}, or large-scale shear flows~\cite{zhou22} -- whose free energy is converted into magnetic energy when small magnetic-field fluctuations whose wavevector is oriented in the colder direction reinforce fluctuations in current density oriented in the hotter direction~\cite{fried59}. The Weibel modes are typically purely growing, filamentary structures whose fastest-growing wavevector is oriented in the direction of the smallest component of the pressure tensor. In general, the electron Weibel is faster than the ion Weibel, because thermal electrons traverse their skin depth much faster than do thermal ions (by a factor ${\sim}m_{\rm i}/m_{\rm e}$). In the limit of weak electron pressure anisotropy, $\Delta T_{\rm e}/T_{\rm e}\ll 1$, the maximum growth rate of the Weibel, $\gamma \sim |\Delta T_{\rm e}/T_{\rm e}|^{3/2} (v_{\rm th,e}/d_{\rm e})$, occurs at wavenumber $k\sim|\Delta T_{\rm e}/T_{\rm e}|^{1/2} \,d^{-1}_{\rm e}$, where $d_{\rm e}\doteq (m_{\rm e}c^2/4\pi n_{\rm e} e^2)^{1/2}$ is the electron skin depth~\cite{weibel59,davidson72}. (Note: $d_{\rm e}\approx 2~{\rm ppc}$ for $n_{\rm e}= 10^{-2}~{\rm cm}^{-3}$.) Recently published analytical and numerical calculations of the Weibel instability occurring in a plasma in which the electron temperature anisotropy is persistently driven by a large-scale shear flow~\cite{zhou22} predict that Weibel can produce ${\sim}0.1~{\rm nG}$ fields in the early ICM. Despite their initially small scales, it is argued that the saturated Weibel seed fields, whose morphology is that of flux ropes, can inverse-cascade to larger scales through magnetic reconnection. While an attractive possibility, this remains to be demonstrated.

Despite our focus here on two plasma processes that provide {\em in situ} generation of cosmological seed fields, it is worth noting an alternative picture in which magnetic fields generated in protogalaxies and/or the first stars were injected into the intergalactic medium or early ICM through winds or outflows (e.g., \cite{rs68,rees87,rephaeli88,fl01}) -- a conjecture supported indirectly by the observed early enrichment of galaxy clusters by metals. For example, a recent paper~\cite{mantz20} uses XMM-Newton observations of a $M \sim 2\times 10^{14}~{\rm M}_\odot$ cluster at redshift $z \simeq 1.7$ that reveal metal enrichment of ${\sim}1/3$ Solar to argue that cluster metallicities were already mostly set when the Universe was less than ${\sim}4~{\rm Gyr}$ old. If such pollutants were to have been accompanied by ${\sim}\mu{\rm G}$ galactic magnetic fields, it is possible that {\em in situ} seed-field generation is unnecessary, depending of course on the efficiency with which such fields were dispersed and diluted throughout, say, a ${\sim}{\rm Mpc}$ cluster volume.

\subsubsection{Plasma theory basics for turbulent dynamo}\label{sec:dynamo_basics}

Now equipped with options for a seed field, how does one amplify it to dynamical strengths? At this point, most physics-focused expositions on the dynamo begin with a collection of statements declaring what is {\em not} a dynamo. These {\em anti-dynamo theorems} all involve some constraints on the allowed symmetries of the velocity fields that can act as dynamos or of the magnetic fields that can be generated by dynamo action. Their essence can be summarized in three words: ``symmetry is bad''.\footnote{There is another type of constraint on dynamo action concerning the minimum magnetic Reynolds number ${\rm Rm}$ below which field amplification is undermined by resistive decay (e.g.,~\cite{rs81,haugen04}), but such a constraint is not at all an issue in the highly conducting ICM.} Here we provide a proof of Zel'dovich's anti-dynamo theorem, most applicable in the ICM in which the dominant motions are random rather than rotational.

Suppose we have a planar flow field $\bb{u}=u_x(t,x,y,z)\ex+u_y(t,x,y,z)\ey$, with $\grad\bcdot\bb{u}=0$ but otherwise arbitrary. The $z$-component of the resistive-MHD induction equation with constant resistivity $\eta$ is simply
\begin{equation}\label{eqn:indz}
    \pD{t}{B_z} + \bb{u}\bcdot\grad B_z = \eta\nabla^2 B_z.
\end{equation}
Multiplying equation~\eqref{eqn:indz} by $2B_z$ and integrating over the volume of the plasma -- a procedure we denote by $\langle\,\cdot\,\rangle$ -- one finds that
\begin{equation}
    \pD{t}{\langle B^2_z\rangle} = -2\eta\langle|\grad B_z|^2\rangle ,
\end{equation}
and so $B_z$ decays resistively to zero. If $B_z=0$, then the solenoidality constraint on the magnetic field, $\grad\bcdot\bb{B}=0$, becomes $\partial B_x/\partial x + \partial B_y/\partial y = 0$. The planar components of the magnetic field may thus be written in terms of a scalar potential, $\bb{B}=\grad\btimes( A\ez)$, which satisfies the un-curled induction equation
\begin{equation}
    \pD{t}{A} + \bb{u}\bcdot\grad A = \eta \nabla^2 A \quad\Longrightarrow\quad \Deriv{t}{\langle A^2\rangle} = -2\eta\langle|\grad A|^2\rangle .
\end{equation}
Again, the magnetic field decays resistively to zero. Thus, no dynamo can be maintained by a planar flow~\cite{zeldovich57}.

What, then, does a fully three-dimensional flow do to a seed magnetic field? In many ways, this question is less about the flow and the field than about the material properties of the plasma, namely, its viscosity and its resistivity. The simplest way to see that this is indeed the case is by examining the ``kinematic'' limit~\cite{kazantsev68,zeldovich84,ka92}, in which the magnetic field is sufficiently weak that the back-reaction on the flow due to the Lorentz force can be neglected. The problem then reduces to solving the (linear-in-$\bb{B}$!) induction equation
\begin{equation}\label{eqn:indohm}
    \Deriv{t}{\bb{B}} = \bb{B}\bcdot\grad\bb{u} + \eta \nabla^2\bb{B}
\end{equation}
governing the evolution of a nearly frozen-in magnetic field as it is sheared, stretched, and compressed by a prescribed velocity field $\bb{u}$ (which we take to be incompressible). With the properties of that velocity field influenced by the plasma's viscosity, $\kappa_u$, and the small-scale structure of the amplified magnetic field regulated by the plasma's resistivity, $\eta$, it is clear that the ratio ${\rm Pm}\doteq \kappa_u/\eta$ -- the {\em magnetic Prandtl number} -- is important. For the remainder of this subsection, we assume a collisional MHD fluid and take ${\rm Pm}\gg 1$. The latter is certain to be the case in the ICM, with Spitzer values for the viscosity and resistivity giving
\begin{equation}
    {\rm Pm} \approx 2 \times 10^{29} \,\biggl(\frac{T}{10^8~{\rm K}}\biggr)^4 \biggl(\frac{n}{10^{-3}~{\rm cm}^{-3}}\biggr)^{-1} .
\end{equation}
%
%
Any reduction in the viscosity by kinetic micro-instabilities (\S\ref{sec:micro}) is unlikely to overcome these 29 orders of magnitude; complications stemming from these instabilities and the anisotropic nature of the viscosity are discussed briefly in \S\ref{sec:plasmadynamo}.

\begin{figure}[t]
    \includegraphics[width=\textwidth]{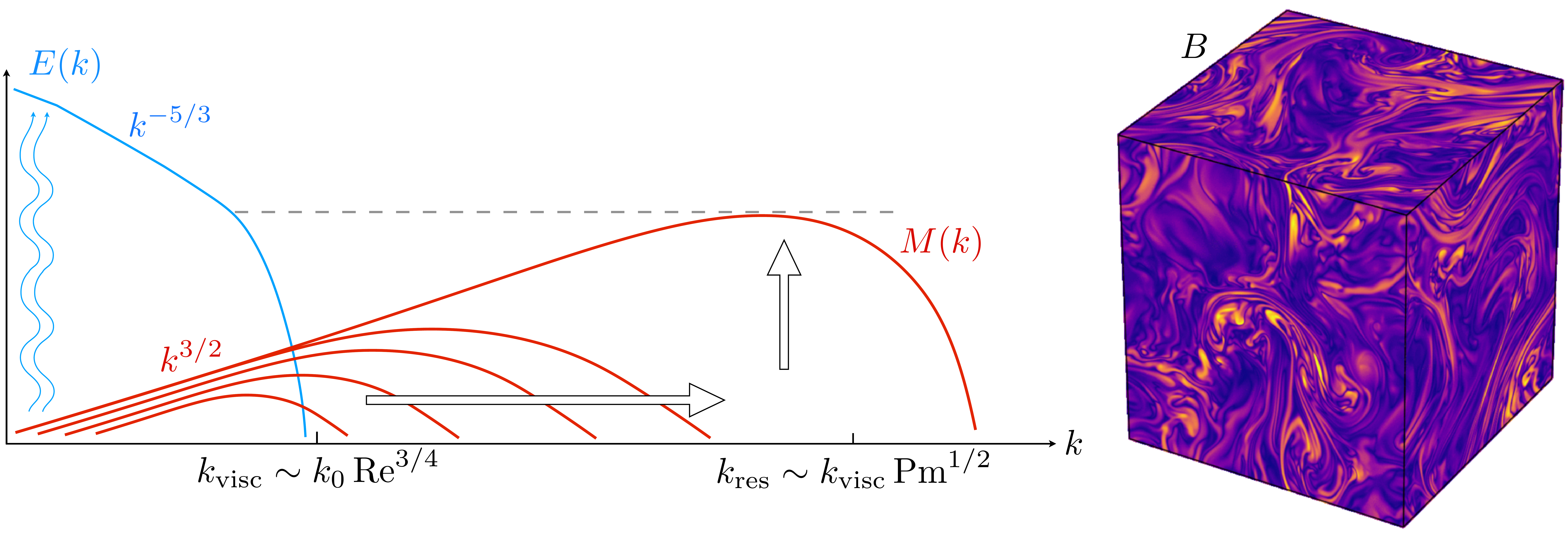}
    \caption{{\bf Left:} A schematic showing the kinetic-energy spectrum $E(k)$ and the evolution of the magnetic-energy spectrum $M(k)$ during the kinematic stage of the turbulent dynamo in an MHD plasma with ${\rm Pm}\gg 1$. Kolmogorov ($k^{-5/3}$) and Kazantsev ($k^{3/2}$) spectral scaling laws are noted. As the viscous-scale motions fold the magnetic field, its energy grows exponentially on small (resistive) scales until it becomes comparable to the kinetic energy of the viscous-scale motions (dashed line) and back-reacts through the Lorentz force. {\bf Right:} A snapshot of the magnetic-field strength $B$ from a three-dimensional direct numerical simulation of MHD turbulent dynamo in a periodic box with ${\rm Pm}=300$ and ${\rm Rm}\sim 10^4$, showing elongated magnetic folds whose energy is concentrated near the resistive scale~\cite{gks22}.}
    \label{fig:dynamo}
\end{figure}

The dynamo then proceeds as follows (see figure~\ref{fig:dynamo}). When ${\rm Pm}\gg 1$, the viscous scale $\ell_{\rm visc}\sim L\,{\rm Re}^{-3/4}$, on which the rate of strain is maximal (see \S\ref{sec:micro}), is much larger than the resistive scale $\ell_{\rm res}$, on which the field's resistive diffusion, $\eta\nabla^2\bb{B}$, becomes competitive with the stretching of the magnetic field by the flow, $\bb{B}\bcdot\grad\bb{u}$ [the two terms on the right-hand side of equation~\eqref{eqn:indohm}]. Using the Kolmogorov scaling $|\grad\bb{u}| \sim (U/L) {\rm Re}^{1/2}$ and setting $\bb{B}\bcdot\grad\bb{u} \sim \eta\nabla^2\bb{B}$ leads to $\ell_{\rm res} \sim L\, {\rm Re}^{-1/4} {\rm Rm}^{-1/2} \sim \ell_{\rm visc}\,{\rm Pm}^{-1/2} \ll \ell_{\rm visc}$. As the viscous-scale motions stretch and fold the magnetic field, its energy spreads over the subviscous range and piles up at the smaller resistive scale, where it grows exponentially at the viscous-scale eddy turnover rate~(e.g., \cite{haugen04}). From the standpoint of these subviscous scales, the action of the flow is roughly equivalent to the application of a random linear shear -- a view that has been leveraged in a variety of analytical models for the kinematic stage of the turbulent dynamo~\cite{kazantsev68,zeldovich84,ka92,scheko02a}, which predict the emergence of an angle-integrated magnetic energy spectrum $M(k) \propto k^{3/2}$, now known as the {\em Kazantsev spectrum}, that is truncated at $k\ell_{\rm res}\sim 1$ by resistive dissipation. The outcome of the kinematic stage is thus a magnetic field organized into elongated folds that reverse direction on the resistive scale (where the magnetic energy peaks), remain fairly straight up to the (viscous) scale of the flow, and exhibit spatial intermittency~\cite{scheko02b,scheko02c,scheko04}. All of these features are evident in the right panel of figure~\ref{fig:dynamo}.

Eventually, the folded field becomes strong enough to back-react on the viscous-scale field-stretching motions via the tension force (when $\langle B^2\rangle \sim U^2\,{\rm Re}^{-1/2}$) and suppress their ability to amplify the magnetic field. Progressively larger (and slower) eddies are then responsible for amplifying the field, while the eddies whose energies are lower, $u^2_\ell \sim U^2 (\ell/L)^{2/3}\lesssim \langle B^2\rangle$, are suppressed. The result is a resistive ``selective decay'' of the magnetic energy at scales too small to be sustainable by the weakened stretching~\cite{scheko02d,scheko04,maron04}. The peak of the magnetic spectrum then shifts towards larger scales while growing secularly in time~\cite{scheko02d,cho09,beresnyak12}, with saturation of the dynamo occurring once the  magnetic energy reaches approximate equipartition with the kinetic energy at the outer scale, $\langle B^2\rangle \sim U^2$. In this state the field and flow self-organize to minimize the parallel rate of strain $\eb\eb\bb{:}\grad\bb{u}$ and thus the stretching of the field~\cite{scheko04,stonge20,seta20}. In low- and intermediate-resolution simulations of the ${\rm Pm}\ge 1$ dynamo, the folded geometry of the magnetic field persists in the saturated state, with a parallel coherence length set by the outer-scale motions and a perpendicular field-reversal scale that remains proportional to the resistive scale as $\ell_{\rm res} \sim L\,{\rm Rm}^{-1/2}$. Recent high-resolution simulations in this regime by Ref.~\cite{gks22} have demonstrated that, at sufficiently large ${\rm Rm}$, such elongated folds are disrupted and broken up by resistive tearing and the magnetic energy spectrum ultimately peaks at a scale that appears to be independent of the resistivity. This last point concerning the emergence of large-scale coherence in the dynamo-generated magnetic field -- hinted at in an earlier simulation~\cite{haugen03} and argued for in some analytic models~(e.g., \cite{subramanian99}) -- is important in the context of galaxy clusters, with Faraday-rotation observations suggesting intracluster field with ${\sim}{\rm kpc}$-scale coherence~\cite{ve05,bonafede10}.

\subsubsection{Enter plasma physics}\label{sec:plasmadynamo}

The vast majority of prior work on turbulent dynamo adopted a fluid (i.e., collisional) MHD framework -- thus the treatment in \S\ref{sec:dynamo_basics}. The physics of weakly collisional, magnetized plasmas such as the ICM are considerably more complicated~\cite{sc06b}. As discussed in \S\S\ref{sec:transport} and \ref{sec:micro}, changes in magnetic-field strength adiabatically produce a field-oriented bias in the thermal motions of the particles -- a bias that induces an anisotropic response in the fluid flow in which field-stretching motions are viscously suppressed. There are two important consequences of this for the dynamo, depending on whether the plasma is collisionless or weakly collisional. First, in a plasma whose particles' magnetic moments are conserved exactly, dynamo amplification of a magnetic field is impossible~\cite{helander16}. An increase in $B$ necessitates a proportional increase in $T_\perp$; the latter must come at the expense of $T_\parallel$. There is simply not enough thermal energy in the plasma to be redistributed in this way if the magnetic field is to increase by more than an order-unity factor. Second, in a weakly collisional plasma subject to field-aligned Braginskii viscosity, it is precisely those motions responsible for growing $B$ ({\em viz.}~$\eb\eb\bb{:}\grad\bb{u}\ne 0$; see equation~\eqref{eqn:lnB}) that are viscously suppressed. With Coulomb collisions in the ICM implying ${\rm Re}_\parallel\sim 1$, this makes for an extremely slow dynamo, with a characteristic timescale determined by the outer-scale motions, ${\sim}L/U \approx 0.5~{\rm Gyr}$ (assuming $L\approx 100~{\rm kpc}$ and $U\approx 200~{\rm km~s}^{-1}$). Fortunately, the pressure anisotropy produced through field amplification drives rapidly growing kinetic instabilities at $\beta\gg{1}$ (as discussed in \S\ref{sec:micro}), which break $\mu$ and effectively increase ${\rm Re}_\parallel$, making the turbulent dynamo possible (and potentially fast) in the ICM.

At the moment there are few simulations of the dynamo in a collisionless or weakly collisional plasma. Ref.~\cite{santoslima14} performed simulations of the turbulent dynamo that evolved the double-adiabatic Chew--Goldberger--Low (``CGL'') fluid equations~\cite{cgl56}, supplemented by an anomalous collisionality intended to mimic the pressure-isotropizing effects of microscale kinetic instabilities. Ref.~\cite{rincon16} used relatively low-resolution hybrid-kinetic (kinetic ions, fluid electrons) simulations to demonstrate that the turbulent dynamo is possible in an initially unmagnetized plasma, and that the dynamo becomes entangled with kinetic-scale dynamical phenomena as the plasma self-magnetizes. Ref.~\cite{stonge18} used hybrid-kinetic particle-in-cell simulations to follow the turbulent dynamo in a collisionless magnetized plasma from its ``kinematic'' stage all the way to saturation, calculating explicitly the effective collisionality of the plasma associated with particle scattering off firehose and mirror instabilities. In Ref.~\cite{stonge20}, semi-analytic modeling and an extensive numerical study using the MHD equations supplemented with a field-parallel viscous (Braginskii) stress were used to shed further light on the turbulent dynamo in a weakly collisional plasma. One important lesson from the latter two studies is that the plasma dynamo is never truly ``kinematic'', in the sense that the parallel viscosity (which sets the maximal rate of strain and thus the amplification rate of the field) knows about the magnetic-field direction and strength ({\em viz.}~${\rm Re} \rightarrow {\rm Re}_{\parallel\rm ,eff}(B)$), even when the latter is energetically insignificant.

As a close to this Section, we speculate on what this field-dependent effective viscosity means for the turbulent dynamo, loosely following some arguments given in Refs~\cite{sc06b,msk16,ms14,stonge20}. If the plasma viscosity were to be dominated by the anomalous contribution from pressure-anisotropy-regulating kinetic micro-instabilities, so that $\kappa_u \sim v_{\rm th,i} \lambda_{\parallel\rm mfp,eff} \sim UL\, (v_{\rm A}/U)^4$ (see equation~(\ref{eqn:predscales}a)), then one may argue that the turbulent dynamo was {\em much} faster in the past, since the induction equation~\eqref{eqn:lnB} then implies
\begin{equation}\label{eqn:lnBeff}
    \Deriv{t}{\ln B} \sim \frac{U}{L} \, {\rm Re}^{1/2}_{\parallel\rm , eff} \sim \frac{U^3}{L} \frac{4\pi m_{\rm i}n}{B^2} \propto B^{-2}.
\end{equation}
Integrating this expression in time and adopting $U\sim 200~{\rm km~s}^{-1}$, $L\sim 100~{\rm kpc}$, and $n\sim 10^{-3}~{\rm cm}^{-3}$ implies that one can achieve ${\sim}\mu{\rm G}$ fields within ${\sim}30~{\rm Myr}$. This is, of course, an attractive possibility from the standpoint of early generation of $\mu{\rm G}$ fields in the ICM. One problem with this scenario, however, is that the enhanced collisionality $\nu_{\rm eff}$ becomes larger than the ion-Larmor frequency -- and is thus unobtainable -- when
\begin{equation}\label{eqn:ultrahighbeta}
    B \lesssim 10\,\biggl(\frac{U}{200~{\rm km~s}^{-1}}\biggr)^{3/5} \biggl(\frac{L}{100~{\rm kpc}}\biggr)^{-1/5} \biggl(\frac{n}{10^{-3}~{\rm cm}^{-3}}\biggr)^{2/5} \biggl(\frac{T}{10^8~{\rm K}}\biggr)^{1/5}~{\rm nG} .
\end{equation}
In this situation~\cite{msk16,stonge20}, the kinetic instabilities do not grow fast enough to efficiently scatter particles and regulate the pressure anisotropy, and the scaling ${\rm Re}_{\parallel,\rm eff} \sim (U/v_{\rm A})^4$ (equation~(\ref{eqn:predscales}b)) should no longer hold. (Exactly what takes its place is an open question.) But as long as ${\sim}10~{\rm nG}$ fields can be produced in the ICM -- either by injection into the intergalactic medium by stellar/galactic outflows or generated {\em in situ} by (slower) dynamo growth -- going from there to the observed ${\sim}\mu{\rm G}$ fields should present little trouble if equation~\eqref{eqn:lnBeff} holds up to further scrutiny.

\section{Energetic particle transport and acceleration in the ICM}
\label{sec:CR}

In addition to the bulk, $\sim$keV thermal plasma associated with signature X-ray emissions, ICM and intergalactic media (hereafter, simply ICM) include populations of non-thermal charged particles with much higher energies, known as cosmic rays (CR). The populations of CR likely include both hadrons (primarily protons, so CRp) and leptons (primarily electrons, so CRe). While these CR are globally sub-dominant in terms of mass and energy content, they still can play important roles in the overall energy balance and dynamics of the ICM. Furthermore, CR emission products, both electromagnetic (radio through $\gamma$-ray) and secondary-particle, can be diagnostic tracers of ICM dynamics and kinetic-scale physics, illuminating structures related to cluster formation and to the interactions between the ICM and embedded cluster components such as individual galaxies, AGNs, and their outflows. These CR-related tracers are increasingly being revealed, sometimes in impressive detail, through observations with an assortment of next-generation instrument technologies. 

Some of these observable CR tracers span cluster-sized domains, demonstrating the importance of cluster-scale CR transport processes. The largest cluster structure-tracing CRe radio synchrotron features indeed span domains that rival those of their home clusters. With the $\mu{\rm G}$-level ICM magnetic fields discussed in \S\ref{sec:obsB}, these radio-visible features require widely distributed populations of $>$GeV CRe often along with active local particle acceleration processes to keep those CRe energized at sufficient energies on Myr timescales, despite unavoidable energy losses, including those associated with their synchrotron emissions.

\begin{figure}[t]
    \centering
    \includegraphics[width=\linewidth]{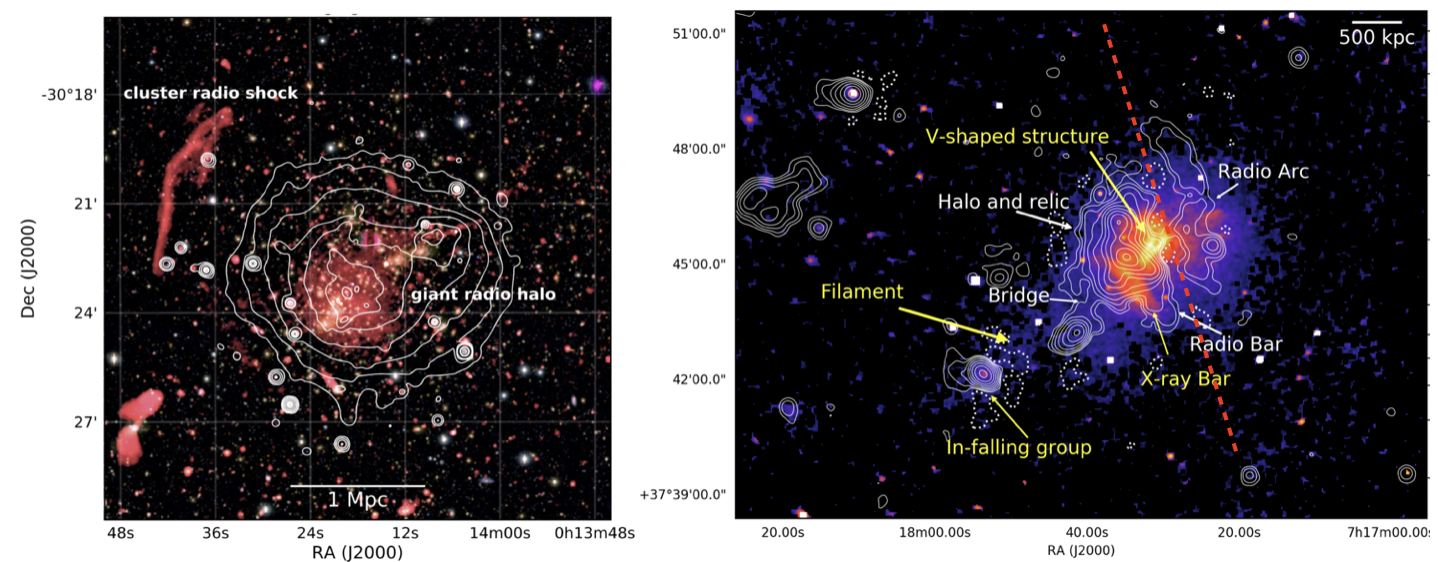}
    \caption{{\bf Left:} Combined radio (VLA 1--4~GHz, red), X-ray ({\it Chandra}, contours), and optical (Subaru, white/yellow) image of A2744. The cluster hosts a giant radio halo and a radio relic/shock. Adapted from Ref.~\cite{vanweeren19}. {\bf Right:} X-ray/{\it Chandra} image of MACSJ0717+3745 with overlaid radio contours (LOFAR). Radio and X-ray emissions trace each other to the left of the red dashed line, while this is not the case to the right. Adapted from Ref.~\cite{bonafede18}.}
    \label{fig:a2744_macs0717}
\end{figure}

\begin{figure}[t]
    \centering
    \includegraphics[width=\linewidth]{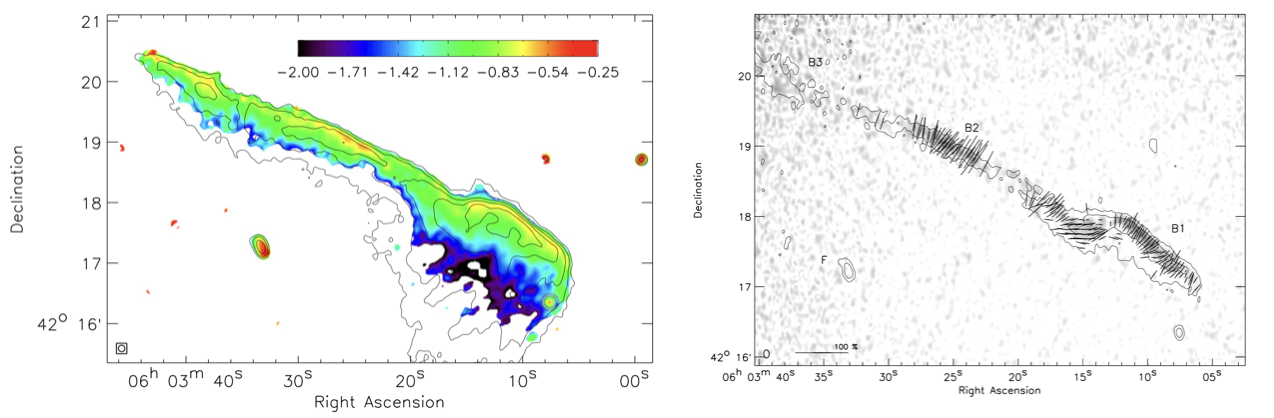}
    \caption{{\bf Left:} Radio spectral index map of the Toothbrush relic based on the measurements between 151 and 610~MHz (color). Contours show the extended radio emission associated with the relic at 151 MHz. Adapted from \cite{vanweeren16}. {\bf Right:} Polarization $E$-vector map of the Toothbrush relic (strokes) measured with the WSRT telescope at 4.9~GHz. The lengths of the strokes represent the polarization fraction; the reference vector for a 100\% polarization fraction is shown in the bottom-left corner. The total polarization is shown as a grayscale image. Adapted from Ref.~\cite{vanweeren12}.}
    \label{fig:toothbrush}
\end{figure}

Cluster-scale radio structures not directly associated with active galaxies are most common in disturbed clusters, such as those involved in mergers. While their detailed properties vary substantially, they are commonly given one of two general labels. When they are centered on the cluster, they are referred to as {\em radio halos} (see figure~\ref{fig:a2744_macs0717} for an example), with those that extend out to cluster extremities being promoted to ``giant halos'' and those that are limited to the core regions being dubbed ``mini halos''. Alternatively, large-scale radio synchrotron features are referred to as {\em radio relics} when they appear as peripheral structures (see figure~\ref{fig:toothbrush} for an example). Such radio relics commonly exhibit elongated structures and the synchrotron emissions are often highly polarized, suggesting reasonably well-ordered magnetic fields in association. Although the origins of these radio relics remain a topic of discussion, the conventional picture associates them with cluster-scale ICM shocks, such as those driven by a merger.

Broadly speaking there are likely to be diverse origins for CR populations in the ICM. A fraction of the CR particles may have been ``injected'' (\ie energized) {\it{in situ}} from the ICM thermal plasma via scattering off ICM-mediated electromagnetic waves generated locally in association with moderately strong ICM shocks (\eg \cite{ha18,ha21}). However, CR injection by such means is likely too inefficient under typical ICM conditions to account directly for most cluster-scale, non-thermal emissions (\eg \cite{petrosian01,vanmarle20,ha21}).  Instead, it appears most likely that the predominant CR populations are more often residual (``fossil'') deposits from galaxy and/or AGN outflows, which have been effectively trapped long term within the ICM by the ICM magnetic fields and then redistributed (and perhaps re-accelerated) through ICM dynamics (\eg \cite{jaffe77,bere97}). There are multiple re-acceleration scenarios that may apply; several are touched upon below. They all share the property that the re-acceleration takes place through a large number of small increments, \ie they are so-called {\em Fermi acceleration} processes. When the increments are all positive, so that the individual CR energies consistently increase, the process is known as ``1$^{\rm st}$ order''; if the increments are stochastic, then the process is ``2$^{\rm nd}$ order''.

As noted at the start of this section, ICM CR are likely to include both CRp and CRe. Unfortunately, at present the ICM lepton/hadron ratios are poorly constrained. The presence of ICM CRp leads to an associated local source for ICM CRe, namely, ``secondary CRe'' by-products of nuclear collisions of CRp with thermal ICM protons. Although it appears that secondary CRe probably are not the dominant CRe species, they still may be an important component, especially in the denser ICM regions (\eg \cite{brun&j14}). 

In what follows we outline briefly some of the underlying physics that governs the spatial transport and energy evolution of CR in the ICM environment. Associated emission processes, as well as the underlying cluster dynamical phenomena that determine the magnetohydrodynamical and thermodynamical state of the ICM, are explored more directly elsewhere in this Chapter and in additional, complementary chapters within this section of the Handbook (``The merger dynamics of the X-ray emitting plasma in clusters of galaxies'' and ``Thermodynamic profiles of galaxy clusters and groups'').

\subsection{Some CR transport basics in the ICM context}

Because CR particles possess very large individual energies, their Coulomb interaction lengths are substantially greater than even those of thermal ICM plasma particles. So, although Coulomb-collision scattering and energy losses are important to the transport of relatively low energy CRe ($E\lesssim 100~{\rm MeV}$)~\cite{sarazin99}, the trajectories and energetics of higher-energy CR are essentially fully colllisionless, being governed instead by Lorentz forces set by local, ambient magnetic-field topologies plus collective, electromagnetic-wave fields of various origins. Particle--wave interactions are addressed below.  For $E\gtrsim 100~{\rm MeV}$, energy losses by CRe are due primarily to synchrotron (S) and inverse-Compton (IC) radiative losses, with IC losses mostly coming from interactions with the CMB radiation field. Both the S and IC energy-loss rates for these CRe scale as $E^2$. Assuming IC losses are predominantly due to CMB radiation, those IC losses at redshift $z$ exceed S losses wherever the local magnetic-field strength $B \lesssim B_{\rm IC}\doteq 3.2\,(1 + z)^2~\mu{\rm G}$. The associated IC energy loss time is $\tau_{\rm IC} \approx 12\,(E/100~{\rm MeV})^{-1}\,(1+z)^{-4}~{\rm Gyr}$. When the local ambient magnetic-field strength is similar to $B_{\rm IC}$, the CRe radiative lifetime from combined IC and S losses is maximal. This translates into a few 100~Myr at small $z$ for CRe energetic enough to emit S radiation in observable radio bands, \ie $E \gtrsim$~a few GeV.

We outline at the end of this section some basics associated with CR transport in association with ICM shocks (including re-acceleration). On the other hand, because the basic transport physics of CR near shocks is mostly similar to CR transport physics more generally in the ICM, we focus first on broader transport behavior. These generally depend on the presence of some level of turbulence, which is ubiquitous in the ICM. The turbulence is observed to be mostly subsonic, although on larger scales it is commonly super-Alfv\'enic, {\em viz.}, with typical fluctuation velocities $\delta v > v_{\rm A}$.

The particle kinetics of energetic CR in the ICM can be expressed in simplified terms as ensembles of particles spiraling along the local magnetic field while undergoing stochastic pitch-angle and momentum scattering by interacting with fluctuating electromagnetic fields associated with ICM-mediated waves and/or instabilities driven by the CR distributions themselves.\footnote{The S and IC emissions do not lead directly to changes in an individual particle's pitch angle, because the momentum carried by emitted photons very nearly aligns with that of the emitting CRe. However, while the loss process does not change the pitch angle of individual CRe, S radiation loss rates are strongly pitch-angle dependent, and so, in the absence of pitch-angle scattering, would lead collectively to the development of an anisotropic CRe distribution biased towards small pitch angles.} Because the relativistic gyroradii of ICM CR are large compared to ICM thermal kinetic scales, the most relevant CR-scattering ICM waves are currently believed to be MHD-scale fast, intermediate (Alfv\'en), and slow modes, whose phase velocities, $v_{\rm ph} = \omega/k$, are set by the thermal acoustic (sound) and Alfv\'en speeds along with the propagation angle with respect to the local ambient magnetic field, $\alpha$. In the high-$\beta$ environment of the ICM, the Alfv\'en speed, $v_{\rm A}$, is much smaller than the adiabatic sound speed, $c_{\rm s}$, while the fast-mode phase velocity, $v_{\rm ph,f} \le c_{\rm s}(1+v_{\rm A}^2/c_{\rm s}^2)^{1/2} \sim c_{\rm s} \ll c$. The Alfv\'en and slow-mode phase velocities, $v_{\rm ph,(i,s)} \le v_{\rm A}$ are much smaller, still. 

At the same time, amplitudes of the most significant scattering-wave magnetic fields, $|\delta B| \sim (\delta v/v_{\rm ph}) B_0$, are for the most part relatively small compared to the ambient mean magnetic field, $B_0$, validating the so-called ``quasilinear'' treatment of CR trajectories within the wave fields. From Faraday's Law, the wave electric and magnetic fields are orthogonal to each other, with $\delta \bb{B} = (c/v_{\rm ph}) \ek\btimes \delta\bb{E}$, where $\ek = \bb{k}/k$.  Consequently, $|\delta\bb{E}| \ll |\delta\bb{B}|$. Since only $\delta E$ can contribute to re-acceleration, this relationship tells us that stochastic pitch-angle scattering by ICM waves, which can come from both wave fields, is generally much faster than stochastic re-acceleration.  For intermediate, Alfv\'en waves, $\delta\bb{E}$ is orthogonal to $\bb{B}_0$ as well as to $\ek$, while for obliquely propagating fast and slow waves components of $\delta\bb{E}$ along $\ek$ are allowed; \ie those modes are partially longitudinal. This is significant to CR scattering by fast mode waves, especially.

Relevant wave sources can be ``extrinsic'', such as cascading ICM turbulence, or ``intrinsic'', such as  waves generated by streaming CR (e.g, in association with a local shock).\footnote{Intrinsic whistler waves near shocks can be important to the transport of relatively low-energy CR near plasma shocks. The behavior of CR at plasma shocks and near sites of magnetic reconnection involves, additionally, consequences of particle trapping within a converging flow.} Because they are the most widely distributed in the ICM, we focus now on extrinsic waves generated by turbulent cascades.

Pitch-angle scattering by these extrinsic waves, which is typically somewhat fast compared to trajectory variations associated with inhomogeneous ambient ICM field structures, strongly drives the velocity-space distribution of CR towards being isotropic.  Being so fast, such scattering can largely counter-balance processes such as those associated with adiabatic particle propagation into an ambient magnetic field pinch or rarefaction that biases towards non-isotropic distributions in pitch angle. Although net equilibrium CR distributions  sometimes exhibit streaming motions (coming from various residual biases towards ``forward'' or ``backward'' pitch angles) or pressure anisotropies (representing symmetric biases towards either large or small pitch angles), such biases can trigger fast-acting CR particle distribution instabilities, such as the firehose or mirror instabilities (\S\ref{sec:micro}), generating intrinsic, CR-mediated waves that strongly  enhance pitch angle scattering to maintain minimal, marginally unstable pitch-angle  anisotropy. Consequently, net streaming motions mostly remain non-relativistic, and pressure anisotropies  also remain small. 

Because stochastic pitch-angle scattering also alters orbital guiding-center motions relative to unperturbed orbits, it inherently leads to spatial particle diffusion. This diffusion can occur both parallel and transverse to the local magnetic field (\eg \cite{desiati14,maiti21}), with distinctively different behaviors. However, the rates of scattering-driven, trans-field CR diffusive transport in ICM contexts are typically slower than the associated CR diffusive transport rates parallel to the local field, and so they are commonly neglected. The rate of spatial diffusion along the local field is controlled by the pitch-angle scattering mean free path, $\lambda$, and is expressed in terms of the spatial diffusion coefficient, $\kappa_{\parallel}$, given, for example by Ref.~\cite{schlick02}, as
\begin{equation}
\kappa_{\parallel} = \frac{\lambda c}{3} = \frac{c^2}{8} \int_0^1 \rmd\pangle \,\frac{(1 - \pangle^2)^2}{\int_0^{\tau}\rmd s\, \langle \dot{\pangle}(0) \dot{\pangle}(s)\rangle},
\label{eqn:kappa}
\end{equation}
where $\pangle \doteq \cos{\theta}$ and $\theta$ is pitch angle, with $\dot{\pangle} = \rmd\pangle/\rmd t$. The time parameter, $\tau$, in the denominator of equation~\eqref{eqn:kappa} measures the correlation time of wave-field variations, while the angle brackets indicate an ensemble average at a given offset, $s$.

There are numerous approaches in the literature evaluating $\lambda$, including quantitative calculations based on detailed scattering treatments for individual wave distributions (\eg \cite{schlick02,brun&j14}). On the other hand, recognizing the complexities and uncertainties that can be inherent in realistic scenarios, other models have approached the problem semi-qualitatively or even heuristically (\eg \cite{miniati15,hopkins21,girichidis21}), including, when non-resonant scattering is important, relating $\lambda$ to a fiducial ICM turbulence-related length (\eg \cite{brun&laz07,miniati15,ptuskin88}). 

Another behavior in the turbulent ICM that may play a key role in CR transport is turbulent magnetic reconnection, which should be ubiquitous (\eg \cite{laz99,kowal20}). Such reconnection has been treated in multiple transport formalisms. Unlike reconnection in low-$\beta$ contexts, such as solar flares, the energy transfer in high-$\beta$ ICM reconnection relates to bulk ICM plasma kinetic energy rather than the explicit dissipation of magnetic field energy. We touch below on one proposed ICM CR transport model that incorporates turbulent magnetic reconnection as a key component \cite{brun&laz16}.

\subsection{Evolution of the ICM CR distributions}

To set up a practical CR transport formalism in the ICM, we focus briefly on characterizing  perturbations to CR trajectories in the ambient magnetic field caused by fluctuating electromagnetic wave fields, applying the common quasi-linear test-particle approach that tracks modifications to unperturbed particle trajectories by linearized waves.  Given the inhomogeneous and turbulent nature of the ICM, the scattering wave fields have limited coherence lengths and times. Consequently, CR orbit perturbations may be dominated by wave-field fluctuation resonances ``selected by the CR'' in their unperturbed particle orbits. These resonance conditions are usually expressed in terms of the Doppler-shifted wave frequency, $\omega^{\prime}$, set by the CR guiding-center motion matching some harmonic of the CR gyro-frequency, $ n\Omega$, where $n$ is an integer. The condition is written as $\omega^{\prime} \approx \omega - k_{\parallel} v_{\parallel} = k_{\parallel}(v_{\rm ph,\parallel} - v_{\parallel}) = n \Omega$, with $v_{\rm ph,\parallel} = \omega/k_{\parallel}$, the wave phase velocity along the local magnetic field, while $v_{\parallel} = v\pangle \approx c\pangle$ is the CR velocity along the local magnetic field.

The strongest scattering is generally associated with either the $n = 0$ or the $n = \pm 1$ resonances. The $n = 0$ resonance ($\omega^{\prime} = 0$) corresponds to the particle's motion along the local magnetic field matching the wave phase speed along the field; that is, the particle ``surfs the wave'', which, given that $v_{\rm ph} \ll c$, primarily restricts the interactions to CR with pitch angles close to 90$^{\circ}$, so that $|\pangle| \ll 1$. Note that this constraint is much more restrictive for the $n = 0$ resonance with the ICM slow mode than with the fast mode.\footnote{The Alfv\'en mode, being purely transverse, does not contribute to TTD interactions (at least linearly).} This resonance is known as  the ``Landau  resonance'' or, in the astrophysics literature, as the transit time damping (TTD) condition.  Imagining the TTD condition, so Doppler shifting the wave frequency to zero in the guiding center frame, it is easy to see that TTD interactions favor a longitudinal $\delta\bb{E}$ field component, \ie one aligned with $\ek$.  The fast mode in this context, being a modified acoustic wave, and thus largely longitudinal and with a larger $v_{\rm ph}$, plays a dominant role in transport models involving TTD.

The $|n| = 1$ resonance is known as the ``gyroresonance condition''. Since $v_{\rm ph} \ll c$, the resonance can be seen to select wave oscillations with wavelength, $2\pi/k \sim 2\pi c/\Omega = 2\pi \rho_{\rm CR}$, with $\rho_{\rm CR}$ the gyroradius of the interacting CR.  In this case, a transverse wave mode, especially one propagating with $|\ek\bcdot\eb_0| = |\cos\alpha| \sim 1$ can be an effective scattering mode. Accordingly, the Alfv\'en mode was traditionally the favored MHD mode in CR transport treatments, especially since it is thought to carry the largest energy flux in MHD turbulence cascades. But, once it was recognized that on small scales relevant to gyroresonances the Alfv\'en mode cascade becomes highly anisotropic, being dominated by waves with $|\cos\alpha| \ll 1$~\cite{gold97}, it has been understood that the Alfv\'en mode is most likely a subdominant contributor to global CR transport (\eg \cite{yan02}). 

Noting that only a wave's $\delta\bb{E}$ field can exchange energy with a CR, it is apparent that particle acceleration per-se by stochastic wave interactions depends on the $\delta\bb{E}$ fields rather than the $\delta\bb{B}$ fields. In particular, when a wave $\delta\bb{E}$ field aligns with and is phased with a CR's  unperturbed motion, the electric field can transfer energy between the wave and the CR for a significant time, thus energizing the CR at the expense of the wave, or the other way around. Either interaction can happen, with the transfer sense dependent on the (random) phase relationship between particle and wave field. The resulting acceleration is stochastic, so a 2$^{\rm nd}$-order Fermi process. 

To apply particle kinetic processes such as those mentioned above to the evolution of ICM CR populations in space and energy (or space and momentum) we need an appropriate transport equation. As justified above we assume the CRe and CRp particle phase-space distributions are nearly isotropic and can be written as $f(p,\bb{x},t)$ at momentum $p$, 3D  spatial coordinate $\bb{x}$, and time $t$. Then, including stochastic scattering interactions leading to spatial and momentum diffusion, the evolution of either the CRp or CRe distribution $f$ can be described by a so-called diffusion-convection transport equation (\eg \cite{schlick02,jk05,brun&j14}),
\begin{align}
\pD{t}{f} &+ \bb{u}_{\rm CR}\bcdot \grad f + \pD{x_\parallel}{}\left(\kappa_{\parallel}\pD{x_\parallel}{f}\right) + \frac{1}{p^3}\pD{y}{}\left[ p^3\left(\frac{D_{pp}}{p^2} \pD{y}{\ln f} - \frac{\dot{p}}{p}\right) f \right]  \nonumber\\*
\mbox{} &= S(p,\bb{x},t)
\label{eqn:fokker}
\end{align}
where $y = \ln{(p)}$, $p = \Gamma \beta$ the particle momentum in units of $m c$, $\Gamma = 1/\sqrt{1 - \beta^2}$, and $\bb{u}_{\rm CR} = \bb{u} + \bb{u}_{\rm s}$ the net CR convection velocity, including any streaming component $\bb{u}_{\rm s}$ with respect to the local thermal plasma flow velocity $\bb{u}$. $\kappa_{\parallel} = (1/3)\lambda c$ is the CR spatial diffusion coefficient along the local mean magnetic field (so, along $x_{\parallel}$; recall that we neglect spatial diffusion transverse to the local magnetic field), and $D_{pp} = p^2/(4 \tau_{\rm a})$ is the CR momentum diffusion coefficient accounting for $2^{\rm nd}$-order Fermi processes, with $\tau_{\rm a}$ the effective stochastic acceleration time. For  all the stochastic acceleration models outlined below, $\tau_{\rm a}$ turns out to be independent of the momentum, $p$. The quantity $\dot{p}$ in equation~\eqref{eqn:fokker} accounts for any monotonic time changes in CR particle momenta, including $1^{\rm st}$-order Fermi processes such as diffusive shock acceleration,  radiative losses, and adiabatic expansion or compression changes, for which $\dot{p}/p = -(1/3)\grad\bcdot\bb{u}_{\rm CR}$. Finally, $S(p,\bb{x},t)$ represents local CR source terms, such as particle injection or escape. The local space density of a given CR component is $n_{\rm CR} = 4\pi \int\rmd p\, p^2 f(p) = \int\rmd p  \, F(p)$. Then, since the relativistic particle energy $E = p c$, the CR volumetric energy spectrum $\rmd n_{\rm CR}/\rmd E = (4\pi/c^3) E^2 f(E/c)$.

\subsection{Some models for $D_{pp}$}

As noted above, a commonly applied model for 2$^{\rm nd}$-order acceleration in the turbulent ICM is based on resonant TTD interactions between CR and isotropic MHD fast-mode waves.  The effective momentum diffusion coefficient, $D_{pp}$, as derived in Ref.~\cite{brun&laz07} can be expressed as
\begin{equation}
D_{pp} = \frac{\pi}{16} \frac{p^2}{c}\frac{W_B}{U_B} \langle k\rangle \int_0^1 \rmd\zeta \left[1 - \left( \frac{v_{\rm ph}}{c\zeta}\right)^2\right] \frac{1 - \pangle^2}{\zeta} \, v_{\rm ph}^2 \, H\biggl[1-\frac{v_{\rm ph}}{c\zeta}\biggr],
\end{equation}
where $W_B = (\delta B)^2/8\pi = \int\rmd k\, W_B(k) $ is the magnetic energy density in fast mode waves, $U_B = B_0^2/8\pi$, $\langle k\rangle = \int\rmd k\, k W_B(k)/W_B$ is the weighted wave number of fast-mode waves, $\zeta \doteq \cos{\alpha} =  k_{\parallel}/k = \ek\bcdot\eb_0$, and $H[\,\cdot\,]$ is the Heaviside step function restricting the integration to include only waves capable of resonant TTD interaction with relativistic CR. Very roughly, the characteristic TTD acceleration time applying this relation is $\tau_{\rm a} \sim (c/v_{\rm ph}) (U_B/W_B)/[\langle k\rangle v_{\rm ph}]$.

As an alternative, and because ICM fast modes are largely longitudinal, compressional waves, they are candidates to accelerate CR through non-resonant stochastic compressions, as formulated in the ISM context by Ref.~\cite{ptuskin88}. Further, as pointed out, for instance, by Refs~\cite{brun&laz07,miniati15} this can, in fact, compare to TTD turbulent re-acceleration in the ICM. If the turbulence fast mode mean square velocity fluctuations cascade in a power spectrum $W(k) = (1/2)m_{\rm i} n [\delta{v}(k)]^2$ for $k \ge k_{\rm min}$ with $k_{\rm min} = 2\pi/L$, and the total mean squared compressive velocity fluctuations are $(\delta v)^2 = \int\rmd k\, [\delta v(k)]^2 W(k)/\int\rmd k\, W(k)$, Ref.~\cite{ptuskin88} showed that $D_{pp}$  from this non-resonant acoustic scattering can be written approximately as
\begin{equation}
D_{pp} \sim p^2 \frac{(\delta v)^2}{c_s^2} \langle k^2\rangle \kappa.
\label{eqn:nonres}
\end{equation}
Here, $\langle k^2 \rangle = \int\rmd k\, k^2 W(k)/\int\rmd k\, W(k)$ and $\kappa = (1/3) c \lambda$ is the spatial diffusion coefficient. The scale $L$ will generally be close to the driving scale for the turbulence of interest. Note, of course, that for solenoidal turbulence most of the turbulent energy in a cascade is carried by Alfv\'en waves rather than fast-mode waves. The importance of fast modes in CR transport comes  to a large extent from the fact that their cascade is expected to remain isotropic to small scales, while both Alfv\'en and slow modes become highly anisotropic with $|\cos{\alpha}| \ll 1$ on small scales. Defining the effective  CR mean free path is complex. Ref.~\cite{brun&laz07} argued that an appropriate relation is $\lambda \sim (1/3){\rm max}[\ell_{\rm diss},{\rm min}(\ell_{\rm A},\lambda_{\rm ICM})]$, where $\ell_{\rm diss}$ is the turbulence dissipation scale, $\ell_{\rm A}$ is the so-called Alfv\'en scale on which the solenoidal turbulence velocity is comparable to the mean field Alfv\'en velocity, while $\lambda_{\rm ICM}$ is the thermal ICM Coulomb scattering mean free path.  Other models assume simply that $\lambda \sim \ell_{\rm A}$ is a suitable estimate in a turbulent, collisionless medium (\eg \cite{brun&laz16}). 

A still different and hybrid transport model combining non-resonant scattering off ICM turbulent magnetic fields combined with $1^{\rm st}$-order acceleration in turbulent magnetic reconnection sites was introduced by Ref.~\cite{brun&laz16}. In this analysis CR gain energy when temporarily trapped in a convergent flow in a reconnection site until they are scattered out of the inflow. As they propagate away from a given site along field lines being stretched as a part of the turbulent dynamo, the CR lose energy in order to conserve the phase space volume they occupied when they escaped the reconnection site. The scattering mean free path outside the reconnection regions is estimated to be comparable to the Alfv\'en scale of the turbulence, $\ell_{\rm A}$, since it is on that scale that ambient field lines become strongly bent due to the turbulent motions. Because the energy gains and losses in this picture are effectively stochastic, the net result is a $2^{\rm nd}$-order acceleration, with the energy step size determined by the local reconnection rate (some fraction of $v_{\rm A}$) combined with the trapping time in a given reconnection site. The effective momentum diffusion coefficient worked out by Ref.~\cite{brun&laz16} is
\begin{equation}
D_{pp} \approx 48 \frac{[\delta v(L)]^3}{c v_{\rm A} L} p^2,
\end{equation}
where $\delta v(L)$ is the amplitude of solenoidal turbulence on the outer scale, $L$. The effective acceleration time, $\tau_{\rm a} \sim (1/200) c v_{\rm A} L/[\delta v(L)]^3$, which can be competitive with analogous acceleration times in the TTD and nonlinear acoustic acceleration pictures outlined above.

One of the key general predictions of the turbulent re-acceleration model is a correlation between spatially resolved variations of turbulence and radio power strength within the halos. While direct velocity measurements of the hot gas in the ICM await future X-ray observatories with high-resolution spectral capabilities (e.g., XRISM, Athena; see \S\ref{sec:future}), this correlation has been checked using an indirect proxy of gas turbulence through the amplitude of density fluctuations \cite{zhuravleva14b}. Ref.~\cite{eckert17}, for instance, found a correlation between the  radio power, $P_{\rm radio}$, at 1.4~GHz and velocity dispersion, $\sigma$, measured on large scales ${\sim}600~{\rm kpc}$, namely, $P_{\rm radio}\propto \sigma^{3.3}$. The proportionality was also established within a cluster MACS0717.5+3745 -- a merging cluster with radio halo emission only partially tracing the thermal X-ray emission of the gas (figure~\ref{fig:a2744_macs0717}, right panel). Using gas density fluctuations as a tracer of gas velocities, Ref.~\cite{bonafede18} found a different ratio of kinetic over thermal energy in the regions with and without radio halo emission (a ${\sim}30$\% difference). While encouraging, the velocities are measured only on large scales in both experiments. These scales are strongly affected by systematics; namely, the choice of an unperturbed structure model. Future direct velocity measurements will verify and refine these results.  

\subsection{CR acceleration in ICM shocks: ``DSA''}\label{sec:DSA}

Acceleration (including re-acceleration) of CR by strong shocks in galactic media are generally thought to be primarily responsible for production of galactic CR up to $\sim$PeV energies. ICM shocks, although being extensive, as well as major vehicles for energy dissipation associated with galaxy cluster formation, are substantially weaker structures than those galactic media shocks. Nonetheless, they are widely invoked to account for acceleration or re-acceleration of CRe to explain the extended and elongated ``radio relic'' S emissions found fairly often in merging clusters (\eg \cite{vanweeren11}; see also figures~\ref{fig:a2744_macs0717} and \ref{fig:toothbrush} for two observed examples). In addition, moderately strong bow shocks seem likely on the advancing heads of some radio galaxies. So, shocks do seem likely do play a significant role in ICM CR budgets.

The acceleration of CR at shocks is customarily described according to the diffusive shock acceleration (DSA) theory 
\cite{bell78a,bo78}. In the classic DSA theory, as they stream near the shock transition in the thermal plasma, energetic particles (with suprathermal to CR energies) resonantly amplify local ICM wave modes (\eg \cite{bell78b,bell04,ha21}). The resulting intrinsic turbulence scatters CR near the shock, temporarily trapping them 
in a converging flow across the shock if their scattering lengths across the shock transition are much greater
than the shock thickness. Trapped particles escape eventually
by convection downstream. Until they do, they gain energy each time they are reflected upstream across the shock,
with a fractional gain determined by the velocity change they encounter across the shock discontinuity; \ie they undergo a $1^{\rm st}$-order Fermi acceleration. 
The net hardness (flatness) of the resulting CR spectrum then reflects the
balance between energy gain and escape rates. In other words, it depends on  
the fractional energy gain in each shock crossing combined with the
probability that particles remain trapped long enough to reach
high energies. This outcome is usually expressed in terms of the sonic Mach number of the shock transition, $M_{\rm s}$, since, ignoring some complicating plasma issues, that controls the velocity jump through the transition. In the absence of local energy losses, the emerging CR distribution is quasi-isotropic with a power-law momentum distribution, $f \propto p^{-s}$, where the index $s$ is given by
\begin{equation}
s = \frac{4 M_{\rm s}^2}{M_{\rm s}^2 - 1}.
\label{eqn:dsa-s}
\end{equation}
As $M_{\rm s} \rightarrow \infty$, $s \rightarrow 4$, while as $M_{\rm s} \rightarrow 1$,  $s \rightarrow \infty$ -- trends that reflect the relatively larger energy boost per shock crossing when the trans-shock velocity jump is larger. Since DSA at ICM shocks is most commonly used to explain distinct radio synchrotron features, and especially radio relics, it is important to note that synchrotron radiation from an electron distribution with a power-law momentum distribution also has a power-law spectral form, with emissivity $\propto \nu^{\alpha}$, where the sign of the spectral index $\alpha = (3-s)/2$ reflects the common sign choice by radio astronomers. From equation~\eqref{eqn:dsa-s} it is easy to show that the so-called ``injection'' synchrotron spectral index, $\alpha_{\rm inj}$, for synchrotron emissions from immediately post-DSA CRe is
\begin{equation}
    \alpha_{\rm inj} = -\frac{M_{\rm s}^2 + 3}{2(M_{\rm s}^2 - 1)} < 0.
    \label{eqn:dsa-alpha}
\end{equation}
Equation~\eqref{eqn:dsa-alpha} can, of course, be inverted to express the associated shock strength in terms of the synchrotron injection spectrum. That is
\begin{equation}
    M_{\rm s}^2 = \frac{2 \alpha_{\rm inj} - 3}{2 \alpha_{\rm inj} + 1}.
    \label{eqn:Mdsa}
\end{equation}
Note that since radio observations of ICM features commonly do not have sufficient angular resolution to isolate immediately post-DSA populations, while S and IC radiative losses will deplete the higher energy CRe faster than the lower energy CRe, steepening the energy spectrum. In a steady state with uniform S and IC loss rates in the post-shock flow, the net effect on the volume-integrated radiation spectral index is to steepen it to $\langle\alpha\rangle = \alpha_{\rm inj} - 1/2$ \eg Ref.~\cite{giacintucci06}.
Then equation~\eqref{eqn:Mdsa} becomes
\begin{equation}
    M_{\rm s}^2 = \frac{\langle\alpha\rangle - 1}{\langle\alpha\rangle + 1}.
\end{equation}

The strong  supernova remnant shocks associated with galactic CR acceleration have $M_{\rm s} \gg 1$, so, as expected, the momentum distribution of galactic CR resembles a power law with index $s \sim 4$ (so that $\alpha_{\rm inj} \sim - 1/2$). Although the energy that drives ICM shocks is enormous, their physical scales are huge, and their sonic Mach numbers, both theoretical and observational, are only a few. So, the expected spectral slopes would be substantially steeper than for many galactic shocks. The character of weak shocks  and their interactions with CR in the high $\beta$ ICM remains something of an open question. For example, some simulations suggest that such shocks with $M_{\rm s} \lesssim 2.3$ may be ``subcritical''; \ie they do not develop the intrinsic turbulence needed to trap low energy particles for acceleration (\eg \cite{kang19,kang21}). Since, as mentioned above, sonic Mach numbers larger than this are likely at least for some merger shocks propagating through the lower density ICM in cluster outskirts, and also for some radio galaxy bow shocks, even with these considerations, DSA seems likely to be a process important to understanding ICM CR.

The most recent LOFAR observations of the Toothbrush relic measured $\alpha_{\rm inj}\approx -0.8$ at the northern edge of the relic and steepening towards the south to $\alpha_{\rm inj}\approx -2$ (see figure~\ref{fig:toothbrush}, left panel). Applying equation~\eqref{eqn:Mdsa} gives a Mach number ${\approx}2.8$ with the lower limit $M\approx 2.5$ -- values that are inconsistent with the Mach number obtained from X-ray observations. This discrepancy between the X-ray- and radio-measured Mach numbers is not unique to the Toothbrush relic and requires additional theoretical work on the DSA scenario for shocks as relic sources in clusters. It is also possible that derived Mach numbers from observations are biased by the complexity of the shock surface, variations of the Mach number along the shock interface, and consequent variations of the shock acceleration efficiency. The relic may also trace multiple shock surfaces along the line of sight. For a complete review and other examples, we refer the reader to the recent review by Ref.~\cite{vanweeren19}.

\section{Future perspectives}\label{sec:future}

In this chapter, we have provided some minimal theoretical background on the physics of weakly collisional, high-$\beta$ plasmas that comprise the ICM and a few observational examples that motivate the importance of this physics. Mostly theoretical and numerical studies have been leading the field in the past decade, establishing a theoretical basis, with {\it Chandra} X-ray observations providing a few tantalizing constraints on transport properties of the ICM and, in combination with radio observations, mechanisms of particle acceleration. Future X-ray missions will offer new observational opportunities by probing smaller-scale physics with more favorable exposure times and higher spectral resolution, significantly advancing observational plasma physics. We close this chapter with a brief review of upcoming or proposed X-ray missions that will be important for the studying the plasma physics of the ICM. 

{\bf X-ray Imaging and Spectroscopy Mission (XRISM)} is a JAXA--NASA collaborative mission with ESA participation, scheduled for  launch in early 2023~\cite{xrism}. The X-ray microcalorimeter on board the mission will offer non-dispersive, high-resolution X-ray spectroscopy with a resolution of ${\sim}5$--$7~{\rm eV}$, allowing direct measurements of bulk and turbulent gas motions in the ICM of many galaxy clusters for the first time. These should provide tighter constraints on the effective viscosity of the ICM in several ways. First, we will be able to calibrate observationally the relation between the amplitude of density fluctuations and the turbulent velocity, which was used previously to constrain the effective viscosity of the bulk ICM in Coma using {\it Chandra} measurements of fluctuations on scales down to those comparable to the Coulomb-collisional mean free path (\S\ref{sec:obsvisc}). Secondly, XRISM will measure velocities of gas motions inside and outside of prominent cold fronts formed in merging clusters, allowing one to determine the formation scenario of the fronts. This will help to overcome current degeneracies that occur when constraining viscosity and conduction locally around the cold fronts' discontinuities (see Chapter ``The merger dynamics of the X-ray emitting plasma in clusters of galaxies'' within this Section for more). Finally, one will be able to measure possible enhancements of gas motions and mixing of metals in the ram-pressure-stripped tails of infalling galaxies (or groups) into the ICM, directly probing the effective gas viscosity. By performing mapping observations of velocities within regions with detected synchrotron emission, one may also probe the turbulent acceleration mechanism of relativistic particles by searching for possible correlations between radio power and turbulence amplitude.

Several mission concepts have been proposed for late 20s to early 30s, including {\bf Advanced Imaging X-ray Satellite (AXIS)}\footnote{https://axis.astro.umd.edu} and {\bf Light Element Mapper (LEM)}\footnote{https://lem.physics.wisc.edu}. The {\it AXIS} concept proposes superb angular resolution on- and off-axis and a factor of ${\sim}10$ larger effective area compared to {\it Chandra}. With such a collecting area and stable resolution within the field of view, it will be possible to obtain detailed maps of gas temperature variations along and across cold fronts (supposedly draped by magnetic fields) to measure electron conduction. Gas viscosity could be constrained through the sharpness of the cold fronts around stagnation points and the sizes of eddies developed by Kelvin--Helmholtz instabilities on the sides of cold fronts, the extent of stripped tails of infalling galaxies, and the power spectra of density fluctuations on Coulomb-mean-free-path scales. With {\it AXIS}'s imaging resolution and large effective area we will discover regions of stretched and amplified magnetic fields (so-called plasma depletion layers), revealing the structure of intracluster magnetic fields and, in combination with high-resolution radio data, mapping the distribution of CR electrons. The LEM concept, in contrast, will provide superb spectral resolution ${\sim}1$--$2~{\rm eV}$ combined with a large grasp (field of view times effective area) that is a factor of a few thousand larger than XRISM's. LEM is intended to observe at lower energies, $0.1$--$2~{\rm keV}$, allowing detailed studies of gas dynamics in massive galaxies, groups, and the outskirts of galaxy clusters. With LEM's superb spectral resolution combined with large grasp, it will be possible to decompose spectrally multiple velocity components of cooler gas phases along the line of sight that would otherwise be difficult to access by any other near-future X-ray mission.

The next-generation X-ray Observatories {\bf Athena}\footnote{https://www.the-athena-x-ray-observatory.eu/} (launch 2034) and {\bf Lynx}\footnote{https://www.lynxobservatory.com} (large mission concept study, planned for early 2040s) will be truly revolutionary for probing plasma physics in the ICM. These missions will combine high-resolution imaging and spectral capabilities with a large effective area, allowing one to build on the rich heritage of discoveries from previous X-ray missions. The physics of weakly collisional plasmas will be probed on a broad range of scales, in the central brightest cluster regions and their outskirts, at low and high redshifts. While specific theoretical predictions are yet to be refined, it will be particularly exciting to measure and compare power spectra of density, pressure, temperature fluctuations, and bulk velocities all the way down to scales on which plasma-physical processes are expected to manifest.

\end{document}